\documentclass[useAMS,usenatbib]{mn2e}
\usepackage{graphicx} 
\usepackage{grffile}
\usepackage{longtable}
\title[The VMC Survey. V. First results for Classical Cepheids]{The
  VMC Survey. V. First results for Classical Cepheids\thanks{Based on
    observations made with VISTA at ESO under programme ID 179.B-2003.}}
\author[V. Ripepi et al.]{V. Ripepi$^{1}$\thanks{E-mail:
ripepi@oacn.inaf.it},
M. I. Moretti$^{2,3}$,
M. Marconi$^{1}$, 
G. Clementini$^{2}$,
M-R. L. Cioni$^{4,5}$,
\and
J. B. Marquette$^{6}$, 
L. Girardi$^{7}$,
S. Rubele$^{7}$,
M.A.T. Groenewegen$^{8}$,
R. de Grijs$^{9,10}$,
\and
B.K. Gibson$^{11,12,13}$,
J. M. Oliveira$^{14}$,
J. Th. van Loon$^{14}$,
J. P. Emerson$^{15}$
\\
$^{1}$ INAF-Osservatorio Astronomico di Capodimonte, Via Moiariello
 16, 80131, Naples, Italy \\
$^{2}$ INAF-Osservatorio Astronomico di Bologna, via Ranzani 1,
Bologna, Italy \\ 
$^{3}$ University of Bologna, Department of Astronomy, via Ranzani 1,
40127, Bologna, Italy \\
$^{4}$ University of Hertfordshire, Physics Astronomy and
 Mathematics, Hatfield AL10 9AB \\
$^{5}$ University Observatory Munich, Scheinerstrasse 1,
 81679 Munich, Germany; Research Fellow of the Alexander von Humboldt Foundation\\ 
$^{6}$ UPMC-CNRS, UMR7095, Institut d'Astrophysique de Paris, F-75014, Paris, France\\
$^{7}$ INAF-Osservatorio Astronomico di Padova, Vicolo
dell'Osservatorio 5,  35122 Padova, Italy \\
$^{8}$ Koninklijke Sterrenwacht van Belgi\"e, Ringlaan 3, 1180, Brussel, Belgium\\
$^{9}$ Kavli Institute for Astronomy and Astrophysics, Peking
University, Yi He Yuan Lu 5, Hai Dian District, Beijing 100871, China \\
$^{10}$ Department of Astronomy and Space Science, Kyung Hee
University, Yongin-shi, 449-701, Kyungki-do, Republic of Korea \\
$^{11}$ Jeremiah Horrocks Institute, University of Central Lancashire,
Preston, PR1 2HE\\
$^{12}$ Department of Astronomy \& Physics, Saint Mary's University, Halifax,
Nova Scotia, B3H 3C3, Canada \\
$^{13}$ Monash Centre for Astrophysics, School of Mathematical Sciences,
Monash University, Clayton, VIC, 3800, Australia \\
$^{14}$ Astrophysics Group, Lennard-Jones Laboratories, Keele University, Staffordshire ST5 5BG\\
$^{15}$ Astronomy Unit, School of Physics \& Astronomy, Queen Mary University of London, Mile End Road, London E1
4NS
}
\begin{document}

\date{}

\pagerange{\pageref{firstpage}--\pageref{lastpage}} \pubyear{2002}

\maketitle

\label{firstpage}

\begin{abstract}
  The VISTA Magellanic Cloud (VMC, PI M.-R. L. Cioni) survey is collecting
  deep $K_\mathrm{s}$-band time-series photometry of the pulsating
  variable stars hosted by the system formed by the two Magellanic
  Clouds (MCs) and the Bridge connecting them.  In this paper we
  present the first results for Classical Cepheids, from the VMC
  observations of two fields in the Large Magellanic Cloud (LMC),
  centred on the South Ecliptic Pole and the 30 Doradus star forming
  regions, respectively.  The VMC $K_\mathrm{s}$-band light curves of
  the Cepheids are well sampled (12 epochs) and of excellent precision
  (typical errors of $\sim$ 0.01 mag). We were able to measure
for the first time the $K_\mathrm{s}$ magnitude of the faintest Classical Cepheids
in the LMC ($K_\mathrm{s} \sim$ 17.5 mag), which are mostly pulsating in the First Overtone (FO) 
mode, and to obtain  FO Period--Luminosity ($PL$), Period-Wesenheit ($PW$), and Period--Luminosity--Colour ($PLC$) 
relations, spanning the full period range from 0.25 to 6 day.
Since the longest period Cepheid in our dataset has a variability
  period of  23 day,
we have complemented our sample with literature data  for brighter
F  Cepheids. On this basis we have built a $PL$ relation in the
$K_\mathrm{s}$ band that, for the first time, includes short period $-$ hence low
luminosity $-$ pulsators, and spans the full range from 1.6 to 100
day  in period. We also provide the first ever empirical $PW$
and $PLC$ relations using the $(V-K_\mathrm{s})$ colour and time-series $K_\mathrm{s}$ photometry. The
very small dispersion ($\sim0.07$ mag) of these relations makes them very well suited to
 study the three-dimensional (3D) geometry of the Magellanic system.
 The use of ``direct'' (parallax- and Baade--Wesselink-  based) distance
measurements to both Galactic and LMC Cepheids, allowed 
us to calibrate the zero points of the $PL$, $PW$, and $PLC$ relations obtained in this
paper, and in turn to estimate an absolute distance modulus of 
$(m-M)_0=18.46\pm0.03$ mag for the LMC. This result is in agreement with
most of the latest literature determinations based on Classical Cepheids.

\end{abstract}


\section{Introduction}

The Magellanic Clouds (MCs) represent a benchmark for studies of
stellar populations and galactic evolution  \citep[see e.g.][]{HZ04,HZ09}. The Milky Way-Magellanic Cloud system is the closest example
 of a complex ongoing galaxy interaction \citep[see e.g.][]{putman98,muller04,stanim04,bekki07}. Having a metallicity lower
 than the Galaxy and hosting significantly younger populous clusters
 the MCs are also an important  laboratory for testing the
 theory of stellar evolution \citep[see e.g.][]{bro04,nl12}.
 
Amongst extragalactic systems, the LMC represents  the ``first step" of the
extragalactic distance scale, thus holding a key role  for the definition of the entire cosmic
distance ladder  \citep[see e.g.][and references therein]{w12}.
% 
%Moreover, the Large Magellanic Cloud (LMC) is the first step of the
%extragalactic distance scale. 
Indeed, the absolute calibration of  extragalactic distances obtained
by the Hubble Space Telescope (HST) Key project \citep[][]{f01} and
the Supernovae Ia (SNIa) calibration team
\citep[see e.g.][]{s01}  both rest upon an assumption of  the distance to the LMC,  and on the  adoption of the $V$- and $I$-bands
period-luminosity ($PL$)  relations of
Classical Cepheids in this galaxy. Any systematic
effects in the distance to the LMC and/or in the slope of the Cepheid $PL$
relations are expected to affect the final calibration of the
cosmic distance scale and, in turn, the resulting estimate of the Hubble
constant \citep[see][and references therein]{m09,b10,w12}.

The Classical Cepheid $PL$ relations have been demonstrated by several authors to show
a non negligible dependence on chemical composition \citep[see
e.g.][]{Caputo00,r05,r08,m09,b10,fm11} with the effect being significantly
reduced at near infrared (NIR) wavelenghts \citep[][]{bccm99,Caputo00,mmf05,m10}.
The NIR bands are also less affected by reddening and, in these filters, 
the $PL$ relations show a smaller intrinsic dispersion \citep[see
e.g.][]{mf91,Caputo00} and a much reduced nonlinearity \citep[][]{bccm99,Caputo00,m09} than in the optical range.
Furthermore, pulsation amplitudes are much smaller in the NIR than in the optical bands, 
thus accurate mean magnitudes can be derived from a small
number of phase points along the pulsation cycle.
%
%{\bf Il pezzo che segue sulle RR Lyrae  spezza il discorso e distoglie l'attenzione. In questo paper  non dobbiamo fare un discorso 
%generale sulla bont\`a delle  PLK per qualunque tipo di variabili,  ma concentrarci sulle Cefeidi, quindi io lo toglierei: 
%An independent estimate of the LMC distance and a calibration of the
%Globular Cluster Luminosity Function \citep{dc06} can also be obtained 
%from the RR Lyrae stars. These Pop.II pulsators are known to
%obey to a PL relation in the NIR bands including a non negligible  metallicity term
%($PLZ$ relation). 
% This relation is
%weakly affected by the evolutionary effects, the spread in stellar mass within
%the instability strip, and uncertainties in the reddening corrections
%\citep[see e.g.][and references therein]{long86,dallora04,coppola11}.
%
%SIETE D'ACCORDO? LO TOLGO???}
%
%
NIR observations of Classical Cepheids, as well as of other pulsating
stars (see, e.g.,  \citealt{Moretti12}, hereinafter M12), 
%Moretti et al. 2012, hereinafter Paper V) 
over the whole Magellanic system, including the Bridge connecting
the two Clouds, are one of the key objectives of the {\it VISTA
  near-infrared $YJK_\mathrm{s}$ survey of the Magellanic system} \citep[VMC;
][hereinafter Paper I]{Cioni11}. This ESO public survey is obtaining
deep NIR imaging in the $Y$, $J$ and $K_\mathrm{s}$ filters
of a wide area across the Magellanic system, using the VIRCAM camera
\citep{Dalton_etal06} of the ESO VISTA telescope
\citep{Emerson_etal06}.  The main science goals of VMC are the
determination of the spatially-resolved star-formation history (SFH)
and the definition of the 3D structure of the whole Magellanic
system. The observations are designed to reach $K_\mathrm{s} \sim 20.3$
mag at Signal to Noise ratio (S/N)=10, in order to detect sources encompassing most phases of
stellar evolution: from the main-sequence, to subgiants, upper and
lower red giant branch (RGB) stars, red clump stars, RR Lyrae and
Cepheid variables, asymptotic giant branch (AGB) stars, post-AGB
stars, planetary nebulae (PNe), supernova remnants (SNRs), etc. These
different stellar populations will enable the study of age and
metallicity evolution within the whole MC system.

In this paper we present results for the Classical Cepheids contained
in the first two ``tiles'' completely observed (the whole 12--epoch
time series) by the VMC survey, namely tiles 8\_8 and 6\_6. Some
preliminary results from the analysis of the Classical Cepheids in
these two tiles were published in \citet{Ripepi12}.  Tile 8\_8 is of
particular interest, as it covers the South Ecliptic Pole (hereinafter
SEP) region that the Gaia astrometric satellite (\citealt{Lin96};
\citealt{Lin10}) %; \citealt{Tur12})
will repeatedly observe for calibration purposes at the start of the
mission, just after launch in Spring 2013.  It is a tile
that lies in an uncrowded, peripheral area of the LMC.
Tile 6\_6 is centred instead on the well known  30 Doradus
(hereinafter 30 Dor) star 
forming region. It lies in the central part of the LMC and is a very crowded area.\\

The VMC data for the SEP and 30 Dor Classical Cepheids are presented
in Section 2. The $K_\mathrm{s}$  $PL$, $PW$, and $PLC$ relations
derived from Fundamental (F) and First Overtone (FO) Classical Cepheids in these two LMC regions are
discussed in Sections 3 and 4.  The zero-point calibrations of the $PL$,
$PW$, and $PLC$ relations based on a number of different methods are
presented in Section 5.
%,  along with the resulting estimates of the distance to the LMC.  
Our final estimate of  the distance to the LMC based on the 30 Dor and SEP Classical Cepheids is discussed in  Section 6. 
 Finally, a summary of the main 
 results is presented in Section 7. 

%
%Optical time-series data in the $V_J$ and $I_{cousins}$
%filters  are available here, from the OGLE~III survey.
%
%The 8\_8 and 6\_6 VMC tiles are centerd  in the South Ecliptic Pole (hereinafter, SEP)  field and on the 
%well known 30 Doradus  (hereinafter, 30 Dor) star forming region of the LMC, respectively. 
%The SEP field is particularly interesting
%because this region of the sky will be continuously and repeatedly monitored during the  
% commissioning phase of the Gaia astrometric mission ( just after the
% launch in Spring 2013. 

\begin{figure*}
\includegraphics[scale=.50]{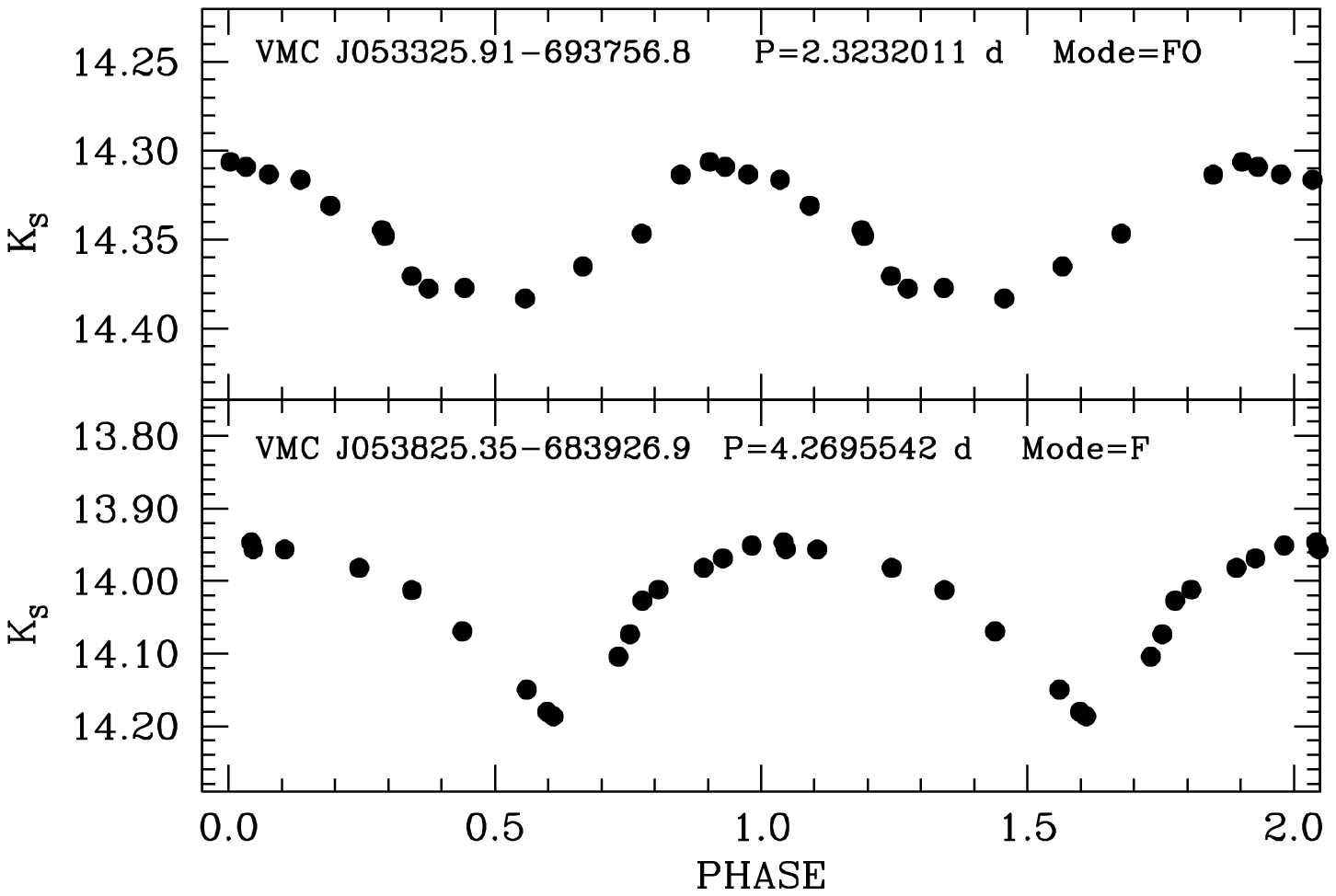}
\includegraphics[scale=.50]{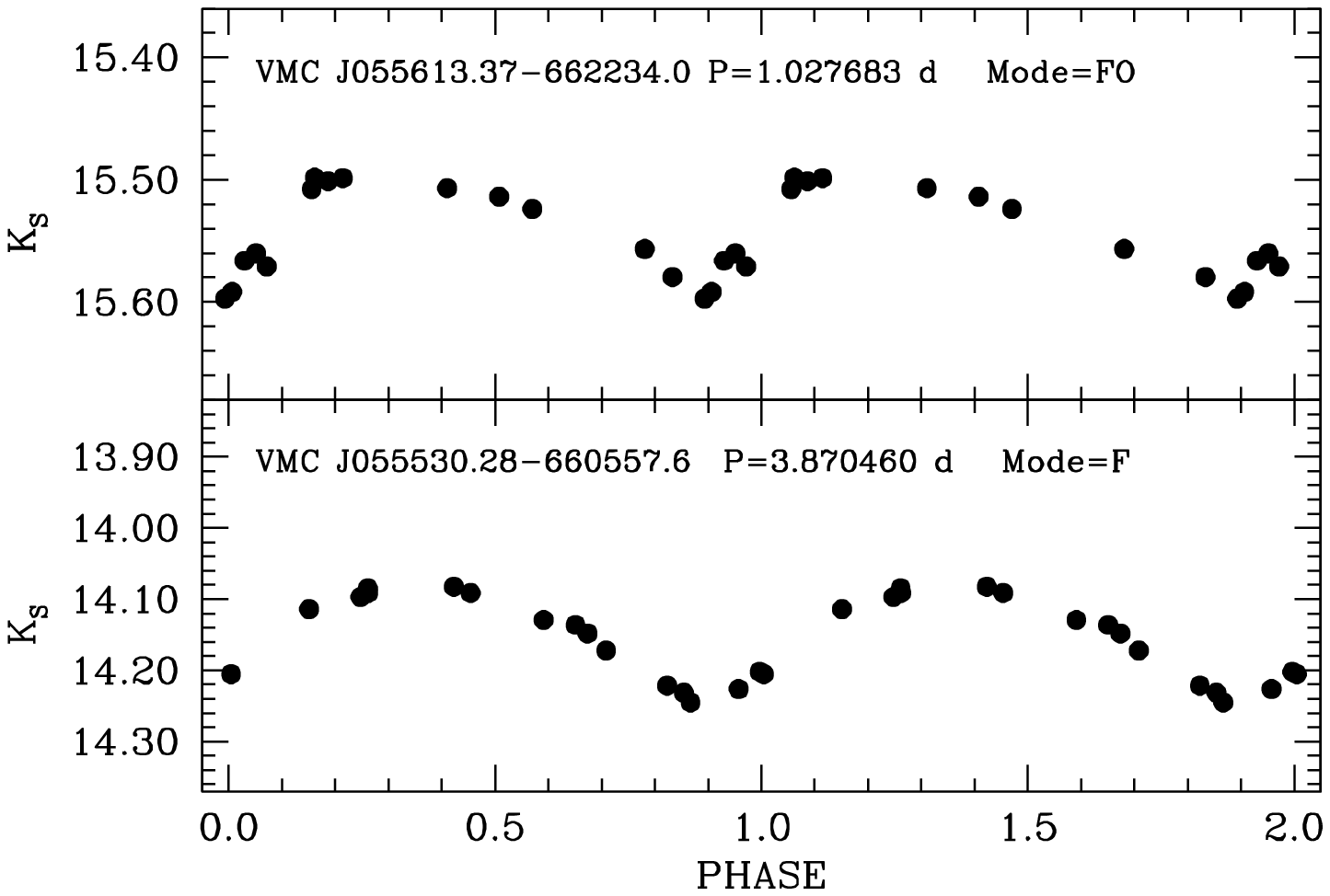}
\caption{Examples of light curves for Cepheids  in the 
30 Dor (left panels) and  SEP (right panels) fields, respectively. Errors of the single-epoch data are of 
the same size as the data points. Periods are from the OGLE-III survey for the 30 Dor Cepheids \citep[][]{sos08,sos09}, and from
the EROS-2 survey for the SEP variables \citep[][]{marque09}. }
\label{figLC}
\end{figure*}

\section{The VMC data for the variable stars}

The VMC observing strategy is described in detail in Paper I. The data
acquisition procedures specifically applied to the variable stars,
the cross-matching between the VMC and existing optical catalogues for
the variable stars, and the derivation of the information needed for
their analysis are extensively discussed in M12.
%
%\citet{Cioni11}  and
The interested reader is referred to these two papers for more
details.  Here, we briefly recall the main steps of the procedures
applied to obtain the $K_\mathrm{s}$ light curves and 
$\langle K_\mathrm{s} \rangle$ average magnitudes for the Classical Cepheids in
the 30 Dor and SEP fields.

In order to obtain well sampled light curves,  
%and in turn measure
%accurate parameters (namely $K_\mathrm{s}$ average magnitudes) for the variable sources, 
the VMC $K_\mathrm{s}$-band time series observations were scheduled
into 12 separate epochs distributed over ideally several consecutive
months.  The VMC data, processed through the pipeline
\citep{Irwin_etal04} of the VISTA Data Flow System
\citep[VDFS,][]{Emerson_etal04}, were retrieved from the VISTA Science
Archive \citep[VSA,][]{Cross12}\footnote{http://horus.roe.ac.uk/vsa/}.
For our analysis we used the v20110909 VMC release ``pawprints'' 
(6 ``pawprints'' form a ``tile'', see Paper I and M12).  Since usually a
variable star is observed in two or three not necessarily consecutive
pawprints, we first calculated a weighted average of the pawprints'
$K_\mathrm{s}$ magnitudes to obtain the ``tile'' $K_\mathrm{s}$, which
then represented one epoch of data.
%\begin{figure}
%\includegraphics[scale=.50]{plotVarGaia.ps}
%\caption{$K_\mathrm{s}$-band light curve for a sample of Cepheids in the SEP
 % field. Note
 %in the third panel the nice light curve for a faint FO Cepheid. Periods were taken form the EROS-2 Survey.}
%\label{fig2}
%\end{figure}
During this process particular care was devoted  to the determination of a proper
Heliocentric  Julian Day (HJD) for the  $K_\mathrm{s}$ value of  each ``tile''  per epoch (see M12, for details). 

The second phase of the ``Exp\'erience pour la Recherche d'Objets
Sombres'' (EROS-2; \citealt{tisserand07}) is, at present, the
  largest optical\footnote {The EROS-2 $blue$ channel (420-720 nm) 
    overlaps with the $V$ and $R$ standard bands, and the $red$
    channel (620-920 nm) roughly matches the mean wavelength of the
    Cousins $I$ band \citep{tisserand07}} survey covering a large
  fraction of the LMC, and reaching out to peripheral areas such as
the SEP region (see Fig. 4 of M12).  The 30 Dor field is covered,
instead, by both the EROS-2 and the third phase of the ``Optical
Gravitational Lensing Experiment'' (OGLE-III;
\citealt[][]{sos08,sos09}) survey.  For our analysis we used
identification, pulsation period, and optical-band light curves from
the EROS-2 photometric archive for the Classical Cepheids contained in
the SEP; however, we opted to use the OGLE-III information, which is
available in standard Johnson-Cousins $V, I$ bands, for the 30 Dor
Cepheids.

%
%
%s of the 
%
%of the Classical  Cepheids contained in the SEP and 30 Dor regions 
%are  available from  the photometric archives of the EROS-2 and OGLE-III Microlensing surveys,  respectively (see Paper V for detaills).
%
%only employed the periods and optical
%(Johnson-Cousins $V, I$ bands) light curves from OGLE-III
%\citep[][]{sos08,sos09}, as the EROS filters are not standard (see above).
%
%The OGLE survey \citep[][and references therein]{sos08,sos09} of which stage IV is in
%progress, covers an area of the MC system that extends progressively  outward
%from the bar of each of the Clouds. EROS-2 \citep{tisserand07} is, at
%present, more extended and covers
%the largest fraction of the VMC field of view including the most peripheral areas such as the SEP 
%region, that in fact  only
%overlaps with EROS-2.  Coordinates, periods and optical\footnote {The 
%EROS-2 $blue$ channel (420-720 nm), overlaps with the $V$ and $R$ standard 
%bands, and the $red$ channel (620-920 nm) roughly matches the mean 
%wavelength of the Cousins $I$ band \citep{tisserand07}} 
%light-curves  of Cepheids in the Gaia SEP field were
%taken from the EROS-2 catalogue and cross-matched to the VMC catalogues for the SEP tile.
%The 30 Dor field is covered by both EROS-2 and OGLE-III,  
%but for the present analysis we only employed the periods and optical
%(Johnson-Cousins $V, I$ bands) light curves from OGLE-III
%\citep[][]{sos08,sos09}, as the EROS filters are not standard (see above). \\

% on the cross-matching between
%the VMC data and the optical catalogues). 

\begin{figure}
\includegraphics[scale=.50]{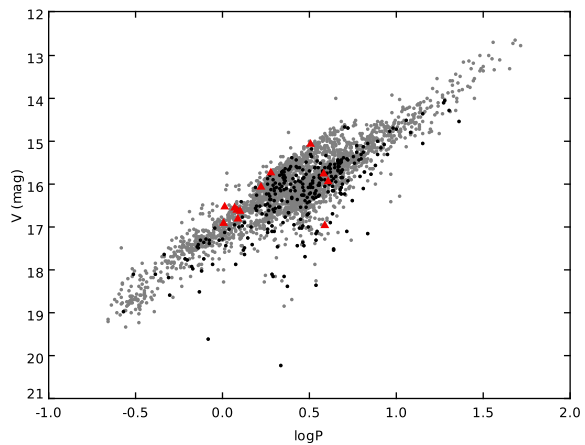}
\caption{Period range covered by the SEP (red filled triangles) and 30
  Dor (black filled circles) Classical Cepheids, over the full range
  of periods spanned by the LMC Classical Cepheids (grey filled
  circles), according to the OGLE-III catalogue
  \citep[][]{sos08,sos09}. In this figure and in the following ones,
  the periods are in day unit.}
\label{ida}
\end{figure}

\begin{table}
\scriptsize
\caption{Sample time-series photometry for the Cepheid VMC J053048.71-694848.0 
 in the 30 Dor field. %Data for all the Classical Cepheids analysed in this paper will be published on-line only.
 }
\label{sampleTimeSeries}
\begin{center}
\begin{tabular}{ccc}
\hline
\noalign{\smallskip} 
HJD-2\,400\,000 & $K_\mathrm{s}$  & err$_{K_\mathrm{s}}$  \\
\noalign{\smallskip}
\hline
\noalign{\smallskip} 
  55140.75594  &    14.347  &    0.007    \\
  55141.77415  &    14.477  &    0.007    \\
  55143.74588  &    14.310  &    0.006    \\
  55147.79060  &    14.394  &    0.006    \\
  55152.80550  &    14.288  &    0.007    \\
  55155.72048  &    14.285  &    0.007    \\
  55161.83663  &    14.336  &    0.007    \\
  55164.77643  &    14.473  &    0.007    \\
  55172.74263  &    14.291  &    0.006    \\
  55191.73701  &    14.280  &    0.006    \\
  55209.66414  &    14.425  &    0.007    \\
  55227.56999  &    14.263  &    0.006    \\
  55246.58263  &    14.275  &    0.006    \\
  55266.51279  &    14.277  &    0.006    \\
  55510.79965  &    14.398  &    0.007    \\
\noalign{\smallskip}
\hline
\noalign{\smallskip}
\end{tabular}
\end{center}
Table~\ref{sampleTimeSeries} is published in its entirety only in the electronic edition of the journal. A portion is shown here for guidance regarding its form and content.
\end{table}

%The OGLE survey \citep[][and references therein]{sos08,sos09} of which stage IV is in
%progress, covers an area of the MC system that extends progressively  outward
%from the bar of each of the Clouds. EROS-2 \citep{tisserand07} is, at
%present, more extended and covers
%the largest fraction of the VMC field of view including the most peripheral areas such as the SEP 
%region, that in fact  only
%overlaps with EROS-2.  Coordinates, periods and optical\footnote {The 
%EROS-2 $blue$ channel (420-720 nm), overlaps with the $V$ and $R$ standard 
%bands, and the $red$ channel (620-920 nm) roughly matches the mean 
%wavelength of the Cousins $I$ band \citep{tisserand07}} 
%light-curves  of Cepheids in the Gaia SEP field were
%taken from the EROS-2 catalogue and cross-matched to the VMC catalogues for the SEP tile.
%The 30 Dor field is covered by both EROS-2 and OGLE-III,  
%but for the present analysis we only employed the periods and optical
%(Johnson-Cousins $V, I$ bands) light curves from OGLE-III
%\citep[][]{sos08,sos09}, as the EROS filters are not standard (see above). \\

As a result of the matching procedure between the VMC and the optical
surveys' catalogues we found 11 Classical Cepheids (of which 8 are FO
and 3 are F pulsators) in the SEP field, and 323 in the 30 Dor region
(of which 161 pulsate in the F mode, 139 pulsate in the FO, whereas 8
and 15 objects are mixed mode F/FO and FO/SO\footnote{SO means Second
  Overtone pulsator}, respectively).  The very small number of
Cepheids in the SEP field may give rise to concerns about the
completeness of the SEP sample.  Indeed, we checked whether new
Cepheids could be identified from the VMC data alone, however 12
epochs do not seem to be sufficient and the variability flag of the
VSA \citep[][]{Cross09} does not yet appear reliable enough for this purpose. On the other
hand, this small number seems to be consistent with the very
peripheral location of the SEP region, which is very far from the LMC
bar, where most of the Classical Cepheids are located.  There are 324
Classical Cepheids in the OGLE-III catalogue of the 30 Dor tile, of
which we recovered 323. Thus in this field we are 99.7\% complete.

Time-series $K_\mathrm{s}$ photometry for these variables is provided
in Table~\ref{sampleTimeSeries}, which is published in its entirety in
the on-line version of the paper.

Our $K_\mathrm{s}$ photometry is in the VISTA system, which  is
  tied to the 2MASS photometry, with the
difference in $K_\mathrm{s}$ magnitude only mildly depending on 
the $(J-K_\mathrm{s})$ colour. Indeed, the empirical results available
to date\footnote{http://casu.ast.cam.ac.uk/surveys-projects/vista/technical/photometric-properties}  
show that: $(J-K_\mathrm{s})$(2MASS)$=1.081(J-K_\mathrm{s})$(VISTA)
and $K_\mathrm{s}$(2MASS)=$K_\mathrm{s}$(VISTA)$-0.011(J-K_\mathrm{s})$(VISTA). 
In the absence of a complete light curve in $J$ (only
very few phase points are available in this passband), this correction might introduce errors larger than
the correction itself. 
Indeed, for the typical $(J-K_\mathrm{s}$) colour of  Classical 
Cepheids with periods shorter than
20--30 d  $(J-K_\mathrm{s} \sim$ 0.3--0.4 mag), the correction is of
on the order of 3-4 mmag,  hence, 
for Cepheids, to a very good approximation, the VISTA system reproduces
well the 2MASS one at $K_\mathrm{s}$. Furthermore, this error for the Cepheids is 
much smaller than the typical uncertainties of the $PL$, $PW$ and $PLC$ relations (see next 
sections).

%
%The light curves are available both
%at the ESO and VSA databases xxx  mettere links xxxx.
The periods available from the EROS-2 and OGLE-III catalogues  
were used to  fold the
$K_\mathrm{s}$-band light curves produced by the VMC
observations of the SEP and  30 Dor variables, respectively.
Examples of  the VMC $K_\mathrm{s}$-band light curves of the Cepheids 
in the 30 Dor and SEP regions are shown in Fig.~\ref{figLC}
 \citep[see also][for additional examples]{Ripepi12}.  
 The light curves are very well sampled and
nicely shaped. Intensity-averaged $\langle K_\mathrm{s} \rangle $ magnitudes were derived from the light curves
%
%
%ence, to derive intensity-averaged
%$\langle K_S \rangle$ we 
simply using custom software written in {\sc C}, that performs
a spline interpolation to the data. Final $\langle K_\mathrm{s}
\rangle$ magnitudes are provided in Table~\ref{tabResults} and
~\ref{tabSep} for the 30 Dor and SEP Classical Cepheids, respectively,
along with the stars main characteristics: VMC Id, coordinates,
pulsation mode, $\langle V \rangle$ and $\langle I \rangle$ (for the
30 Dor Cepheids only) intensity-averaged  magnitudes, period, $K_\mathrm{s}
$-band amplitude
and individual $E(V-I)$ reddenings (for the 30 Dor Cepheids only. See
next section).  The predominance of FO (8) with respect to F
(4) pulsators, as well as
the lack of Cepheids with periods longer than 4 d in the SEP
field are both remarkable. This might be related to the specific star formation history
in the SEP region (see M12 
%Moretti et al. 2012, MNRAS, submitted
for details).

 In the VMC data a significant departure from linearity due to saturation starts
  around $K_\mathrm{s} \sim$ 11.5 mag,  the actual value depending on
  seeing, airmass etc. (see Paper I).  This limits the Cepheids that can be 
  analyzed on the basis of the VMC data to variables with 
 % means that we cannot obtain the 
% photometry  for Cepheids with 
pulsation period shorter than 20--30 day.  The longest period
Classical Cepheids analyzed in the present paper has a variability
period of 23 day.
%In the specific case of the Classical Cepheids analyzed in the present paper, the 
%longest period Cepheid in our sample is a 23 days variable,
However, this threshold was mainly set by the lack of longer period
 pulsators in the OGLE-III and EROS-2 catalogs for the 30 Dor and SEP fields, rather  than by the VMC saturation limit.
%
%30 Dor field, the limit is set 
%  the longest period Cepheid in our sample is a 23 days, 
%due to the lack of longer period pulsators in the OGLE catalog.

Figure~\ref{ida} shows the period range covered by the SEP (red filled
triangles) and 30 Dor (black filled circles) Classical Cepheids, over
the full range of periods spanned by the LMC Cepheids (grey filled
circles), according to the OGLE-III catalogue \citep[][]{sos08}.  The
variables in the SEP and 30 Dor fields appear to sample very well the
full distribution of LMC Cepheids with periods shorter than 20--30
day, thus ensuring a large significance of the $PL$, $PW$ and $PLC$
relations presented in the present study, that were extended beyond
the period limit of 20--30 day set by the saturation threshold of the
VMC $K_\mathrm{s}$ exposures by complementing the SEP and 30 Dor
samples with Classical Cepheids exceeding the 10 d period from
\citet[][see Section 3]{persson}.

%Error bars of the individual $K_\mathrm{s}$ measurements are shown in
%the figures. The uncertainties for the Cepheids %in the 30 Dor field 
%are of the same size at the data-points.

%\input{tabResult}
%\input{tabResultPart}
\begin{table*}
\scriptsize
\caption{Results for Classical Cepheids in the 30 Dor field. M:
    Pulsation Mode. The
  complete table is available in electronic form only. An
``a'' in the last column means that the star was not used to derive the
$PL$, $PW$ and $PLC$ relations.
 }
\label{tabResults}
\begin{center}
\begin{tabular}{ccclccccccccc}
\hline
\noalign{\smallskip} 
ID & RA  & Dec  & M & $\langle I \rangle$ & $\langle V \rangle$ & Period & $\langle K_\mathrm{s} \rangle$ & $A(K_\mathrm{s})$ & $\sigma_{\langle K_\mathrm{s} \rangle}$& $E(V-I)$ & Notes \\
  & J2000  & J2000  &  & mag & mag & d & mag  & mag & mag & mag &   \\
\noalign{\smallskip}
\hline
\noalign{\smallskip} 
VMC J053048.71-694848.0  &   82.70296  &  -69.81333   &   F       &   15.354    &   16.105    &   3.240862         &   14.353     &   0.22     &   0.006     &       0.08      &                   \\
VMC J053059.48-693531.2  &   82.74783  &  -69.59200   &   F       &   14.890     &   15.693    &   4.656827        &   13.917     &   0.22     &   0.008     &       0.09      &                    \\
VMC J053100.91-694532.4  &   82.75379  &  -69.75900   &   F       &   14.724    &   15.453    &   4.874545        &   13.775     &   0.23     &   0.004     &       0.08      &                    \\
VMC J053101.03-690630.3  &   82.75429  &  -69.10842   &   F       &   15.779    &   16.722    &   3.130618         &   14.589     &   0.12     &   0.007     &       0.12      &                    \\
VMC J053101.70-690621.5  &   82.75708  &  -69.10597   &   F       &   15.993    &   17.003    &   2.908321         &   14.694     &   0.20     &   0.010      &       0.12      &                    \\
VMC J053102.99-693207.2  &   82.76246  &  -69.53533   &   F       &   14.816    &   15.547    &   4.641402         &   13.957     &   0.25     &   0.009     &       0.09      &                    \\
VMC J053112.67-700427.3  &   82.80279  &  -70.07425   &   F       &   14.949    &   15.679    &   4.226975         &   14.041     &   0.22     &   0.008     &       0.02      &                    \\
VMC J053117.49-695428.3  &   82.82287  &  -69.90786   &   F       &   14.357    &   15.133    &   5.976499         &   13.388     &   0.21     &   0.004     &       0.10      &                    \\
VMC J053118.30-693626.4  &   82.82625  &  -69.60733   &   F       &   14.875    &   15.716    &   4.83438          &   13.833     &   0.14     &   0.008     &       0.10      &                    \\
VMC J053122.41-695323.0  &   82.84337  &  -69.88972   &   F       &   15.058    &   15.643    &   3.408864         &   14.120      &   0.19     &   0.009     &       0.10      &                    \\
\noalign{\smallskip}
\hline
\noalign{\smallskip}
\end{tabular}
\end{center}
\end{table*}

\begin{table*}
\scriptsize
\caption{Results for Classical Cepheids in the SEP field.}
\label{tabSep}
\begin{center}
\begin{tabular}{ccclccccc}
\hline
\noalign{\smallskip} 
ID & RA  & DEC  & M & $\langle V \rangle$ & Period & $\langle K_\mathrm{s} \rangle$ & $A(K_\mathrm{s})$ & $\sigma_{\langle K_\mathrm{s} \rangle}$ \\
    & J2000  & J2000  &  & mag & d &  mag & mag & mag  \\
\noalign{\smallskip}
\hline
\noalign{\smallskip} 
VMC J055635.76-654742.2  &  89.14900	  &  -65.79506     &  FO   &      16.596   &     1.188733   &  15.264    &  0.09   &  0.0037            \\
VMC J055711.13-655116.1  &  89.29636	  &  -65.85448     &  FO   &      16.561   &     1.044436   &  15.284    &  0.10   &  0.0040	     \\
VMC J055638.33-660302.5  &  89.15971	  &  -66.05070     &  FO   &      16.640   &     1.214786   &  15.338    &  0.08   &  0.0040	     \\
VMC J055530.28-660557.6  &  88.87615	  &  -66.09933     &  F    &      15.785   &     3.870460   &  14.145    &  0.15   &  0.0062	     \\
VMC J055613.37-662234.0  &  89.05570	  &  -66.37611     &  FO   &      16.944   &     1.027683   &  15.533    &  0.10   &  0.0060	     \\
%VMC J055924.79-662930.0  &  89.85330	  &  -66.49167     &  F    &      16.844   &     1.238489   &    15.681    &  0.09   &  0.0113	     \\
VMC J060325.16-663124.5  &  90.85483	  &  -66.52348     &  FO   &      16.677   &     1.277076   &  15.224    &  0.14   &  0.0154	     \\
VMC J060318.77-665244.3  &  90.82822	  &  -66.87896     &  FO   &      15.096   &     3.227865   &  13.631    &  0.09   &  0.0045	     \\
VMC J060117.35-665319.9  &  90.32228	  &  -66.88885     &  F    &      15.986   &     4.085779   &    13.900    &  0.03   &  0.0070	     \\
VMC J055922.13-665709.6  &  89.84220	  &  -66.95267     &  FO   &      16.100   &     1.683674   &  14.788    &  0.11   &  0.0032	     \\
VMC J055535.43-670217.4  &  88.89761	  &  -67.03818     &  F    &      17.009   &     3.902331   &    14.130    &  0.02   &  0.0079	     \\
VMC J055942.93-670346.8  &  89.92889	  &  -67.06300     &  FO   &      15.767   &     1.907595   &  14.515    &  0.09   &  0.0023	     \\
\noalign{\smallskip}
\hline
\noalign{\smallskip}
\end{tabular}
\end{center}
\end{table*}

\section{Classical Cepheids in the 30 Dor Field}

\begin{table*}
\small
\caption{$PL$, $PW$ and $PLC$ relations for F and FO Classical
 Cepheids.  The Wesenheit function is defined as:
 $W(V, K_\mathrm{s})=K_\mathrm{s}-0.13(V-K_\mathrm{s})$.}
\label{tab1}
\begin{center}
\begin{tabular}{cccccccc}
\hline
\noalign{\smallskip}
mode & $\alpha$ & $\sigma_{\alpha}$ & $\beta$ & $\sigma_{\beta}$ & $\gamma$ & $\sigma_{\gamma}$ &r.m.s. \\
\noalign{\smallskip}
\hline
\noalign{\smallskip}
\multicolumn{8}{c}{$K_\mathrm{s}^0$=$\alpha$+$\beta$ log$P$} \\
\noalign{\smallskip}
\hline
\noalign{\smallskip}
F   & 16.070 & 0.017 & -3.295 & 0.018 & & & 0.102  \\
FO & 15.580 & 0.012 & -3.471 & 0.035 & &  & 0.099\\
\noalign{\smallskip}
\hline
\noalign{\smallskip}
\multicolumn{8}{c}{$W(V,K_\mathrm{s})=\alpha+\beta $ log$P$} \\
\noalign{\smallskip}
\hline
\noalign{\smallskip}
F & 15.870 & 0.013 & -3.325 & 0.014 & &  & 0.078\\
FO   & 15.400 & 0.008 & -3.530 & 0.025 & & & 0.070  \\
\noalign{\smallskip}
\hline
\noalign{\smallskip}
\multicolumn{8}{c}{$K_\mathrm{s}^0=\alpha+\beta$ log$P$$+\gamma (V-K_\mathrm{s})_0$} \\
\noalign{\smallskip}
\hline
\noalign{\smallskip}
F   & 15.740 & 0.073 & -3.346 & 0.013 & 0.216& 0.014& 0.073  \\
FO & 15.355 & 0.070 & -3.545 & 0.026 & 0.163& 0.014 & 0.070\\
\noalign{\smallskip}
\hline
\noalign{\smallskip}
\end{tabular}
\end{center}
\end{table*}

\begin{figure}
\includegraphics[width=8cm]{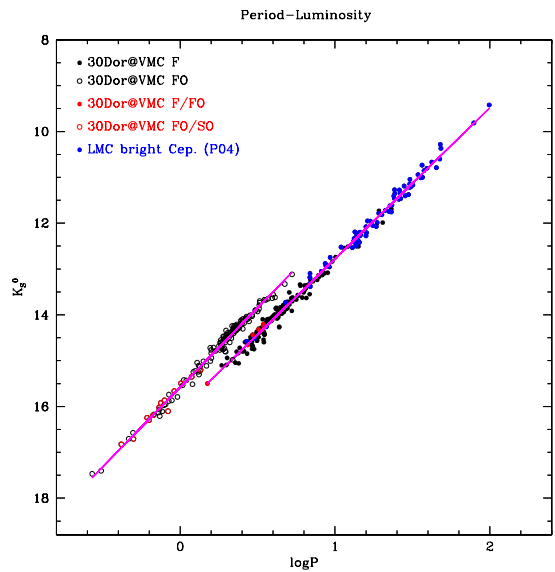}
\caption{$K_\mathrm{s}$-band $PL$ relation for Cepheids in 
the 30 Dor field. Black open and filled circles show FO- and F-mode   
pulsators, respectively. Blue filled circles show the F-mode 
Cepheid sample of \citet{persson}. The solid lines are the result of the least 
square fits to the data (see text for details).} 
\label{figdorCep}
\end{figure}

\begin{figure}
\includegraphics[width=8cm]{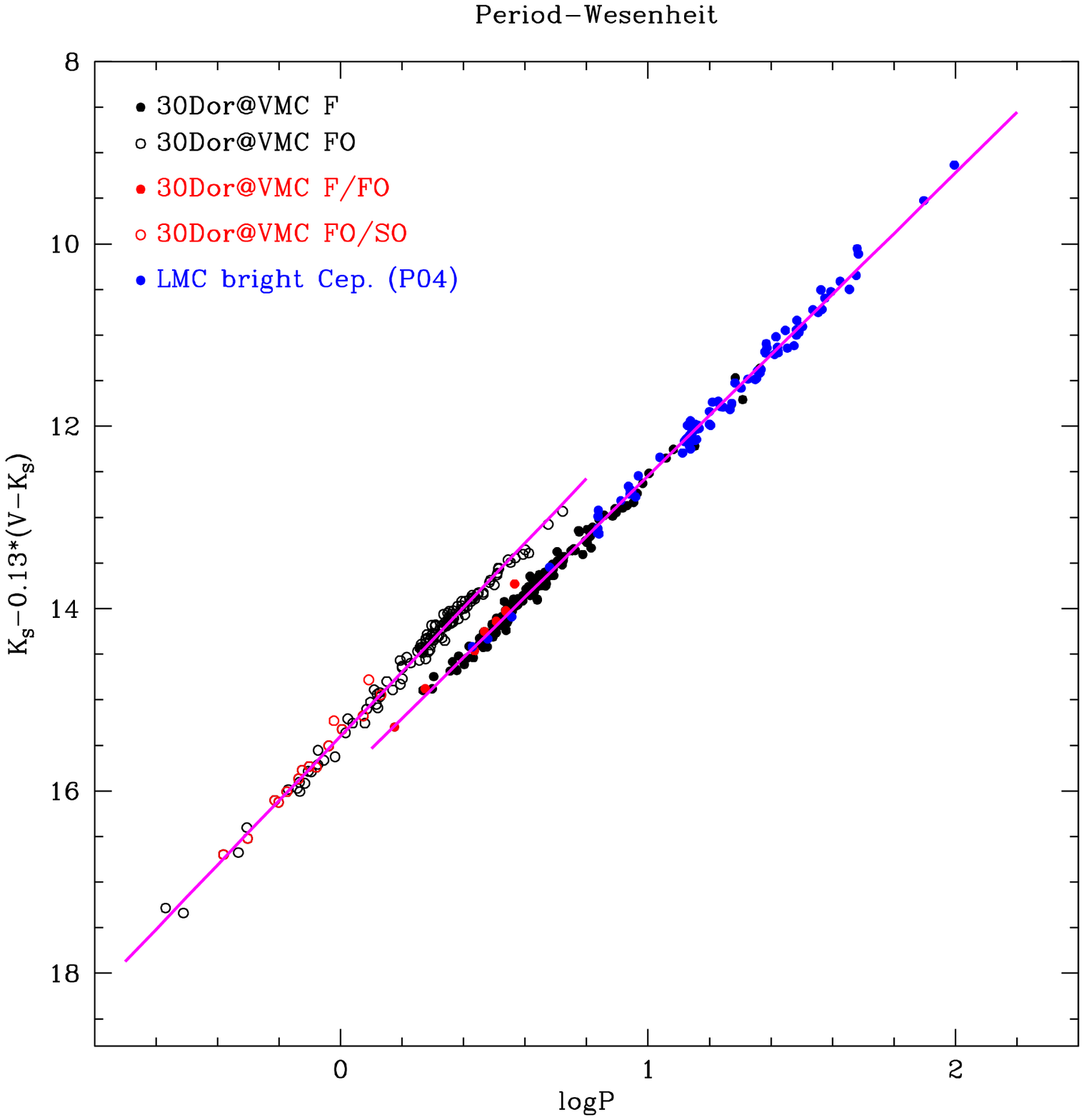}
\caption{$PW$ relation for Cepheids in
the 30 Dor field. Symbols are as in Fig.~\ref{figdorCep}} 
\label{figWas}
\end{figure}

\begin{figure}
\includegraphics[width=8cm]{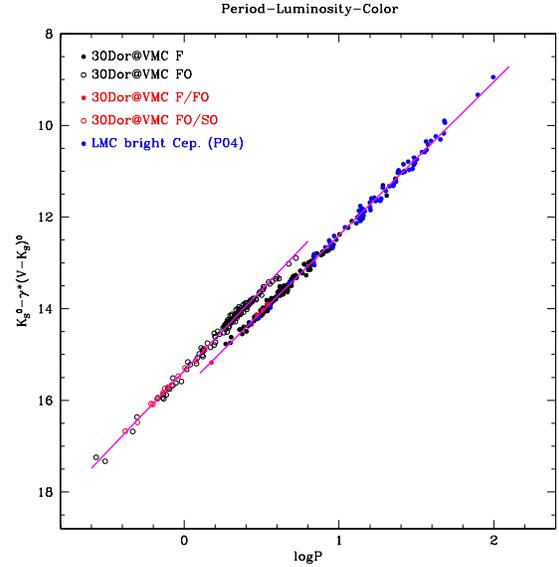}
\caption{$PLC$ relation for Cepheids in
the 30 Dor field. Symbols are as in Fig.~\ref{figdorCep}.} 
\label{figPLC}
\end{figure}

%The 30 Dor field contains 164 F,  139 FO, 8 F/FO, and  15 FO/SO  (SO stands for second overtone) pulsators.
The $PLK$ relations of the 172 F  and 154 FO\footnote{Double mode pulsators F/FO and FO/SO were included
in the F and FO samples, respectively.}
Classical Cepheids in the 30 Dor field are shown in
Fig.~\ref{figdorCep}. 
%{\bf Since the saturation level of the VMC survey in
%$K_\mathrm{s}$ limits the length of the periods we were able to 
%measure to a maximum value of about 15-20 days,}  
In order to extend the period coverage beyond the 
 limit of 23 day set by the longest period pulsator in our sample,
%  Classical Cepheids analyzed in the present paper is a 23 days variable.
%
%Since the longest period included in our sample is around 23 days,}   
we have complemented our
data with the sample of \cite{persson} which includes 84 F-mode
Cepheids with periods mainly ranging between 10 and 100 day. To merge
the two samples we first transformed Persson et al.'s original
photometry from the Las Campanas Observatory (LCO) to the 2MASS system
using the relations of \citet{carpenter}.  These data are shown as
blue filled circles in Fig.~\ref{figdorCep}.  
Inspection of this figure (or equivalently Fig.~\ref{figWas} or
  Fig.~\ref{figPLC}) and the straight line fits to the two sets of
  data shows no obvious discontinuity between the data, indicating
  that our approximation $K_\mathrm{s}$(VISTA)$\approx$$K_\mathrm{s}$(2MASS) 
does not introduce a significant error.

To account for the
variable reddening which characterizes the 30 Dor field, we adopted the
recent evaluations by \citet{haschke} (reported in column 11 of
Table~\ref{tabResults}), while to correct the \citet{persson} dataset we
adopted the reddening values provided by the authors.  We have
verified that the two reddening systems are consistent with each other
within a few hundredths of a mag, and that there is no trend with
period.

Finally, we performed least-squares fits to the data of  F- and
FO-mode variables separately, adopting an equation of the form 
$K_\mathrm{s}^0=\alpha+\beta$ log$P$. The coefficients derived from the fits
are provided in the first portion of Table~\ref{tab1}. 

In addition to  the $PL$ relation in the $K_\mathrm{s}$ band we can consider the $PW$ and 
%Period-Wesenheit (PW) and Period--Luminosity--colour (PLC)
$PLC$ relations. The advantages of  using these relations in place of
a simple $PL$ relation have been widely discussed in the
literature \citep[see
e.g.][]{st68,m82,str09,Caputo00,mmf05,b08,b10,nk05,n12}. %Here we only
%remind that 
These relations include a colour term with a coefficient that, in the
case of the $PLC$ relations, takes into account the colour distribution
of the variable stars within the instability strip, whereas in the
case of the Wesenheit functions it corresponds to the ratio between total
to selective extinction in the filter pair \citep{m82,Caputo00}, thus
making the Wesenheit relations reddening free. We emphasize that these
tools are particularly suited to studying the 3D structure of the
Magellanic system, as they have much smaller dispersions than a 
simple $PL$ relation
\citep{Caputo00,mmf05,b10}. \\
The $PW$ and $PLC$ relations are usually calculated using the $(V-I)$
colour.  However, given the data available to us we have built our
relations using the $V-K_\mathrm{s}$ colour.
%
%we need a colour. Given the data
%available to us we adopted ($V-K_S$). This is not the usual choice form of the PW 
%which is usually calculated using the $(V-I)$ colour. Hence, 
As far as we know, this is the first empirical $PW$ relation using
such a colour.  Following \citet{cardelli89} the Wesenheit function is
defined as $W(V,K_\mathrm{s})=K_\mathrm{s}-0.13(V-K_\mathrm{s})$ which is
correlated with the logarithm of the period according to a linear
relation of the form $W(V,K_\mathrm{s})=\alpha+\beta$ log$P$. Similarly,
we adopt a $PLC$ relation of the form $K_\mathrm{s}^0=\alpha+\beta$
log $P+\gamma (V-K_\mathrm{s})_0$.  The coefficients of the relations
derived with this procedure are provided in the middle and lower
portions of Table~\ref{tab1}. The relations are shown in
Figs.~\ref{figWas} and \ref{figPLC}, respectively.
 
We point out that in deriving the above relations we decided to not apply any 
corrections for the
inclination of the LMC disc to  
%It is important to mention here that to derive the relations above we
%not correct 
\citet{persson}'s  and our 30 Dor Cepheids, since  our attempt 
%when we tried 
 to calculate, 
e.g., the PW relation after 
%having 
de-projecting both Cepheid samples with % by using
the widely adopted  \citet{vandermarel01} or \citet{vandermarel02} model
parameters resulted in an increased  r.m.s. dispersion of the relation. This is  likely because 
%
%probably
%due to the fact that the uncertainties of 
current model  uncertainties introduce
errors of comparable size as the corrections themselves. In our
particular case, as \citet{persson}'s  Cepheids are spatially well distributed
over the whole LMC,  their use should not introduce any systematics on
the distance to the LMC barycentre. On the other hand, according
to the aforementioned LMC disc models, the 30 Dor field is displaced off the LMC 
barycentre by, on average,  only $\sim$0.02 mag \citep[see also Table 3 in][who performed similar
calculations using the comparison between observed and simulated
CMDs]{Rubele12}. Such an effect is  much smaller than the intrinsic
scatter of the PW relations. We therefore believe  that the 
relations in Table~\ref{tab1}  can be safely applied to estimate  the distance
to the LMC barycentre.

We can now compare our results with previous studies.  The literature
values for the coefficients of the $PL$, $PW$, and $PLC$ relations are
summarized in Table~\ref{comparison}.  The first four rows in the
table report the empirical results for the $PLK_\mathrm{s}$ relation of
F pulsators by \citet{groe00,persson,Testa07,Storm11b}, while
the fifth row shows the theoretical results by
\citet{Caputo00}. Similarly, rows 6-7 display the empirical results by
\citet{groe00} and the semi-empirical results by \citet{Bono02}, for FO pulsators.  A
comparison between Tables~\ref{tab1} and~\ref{comparison} reveals
that for the $PLK_\mathrm{s}$ relation of the F-mode pulsators, there
is general agreement within the errors between our results and
\citet{groe00,persson,Storm11b} (only slope for the latter because the
zero point is given in absolute magnitude). Only marginal agreement is found
instead with \citet{Testa07}, whose results are based on the
\citet{persson} sample complemented at shorter periods by Cepheids
belonging to the LMC clusters NGC1866 and NGC2031. This is likely 
due to the significantly larger sample, both at short and medium periods, presented in
the present paper. As for the comparison with theory, we find that there is
a satisfactory agreement between our slope and the slope predicted by
pulsation models for the LMC's chemical composition \citep{Caputo00}.
The errors in the coefficients of our $PLK_\mathrm{s}$ relation are
shorter than in previous studies. This is a result of the large range in
period spanned by the Cepheids in our sample, including for the first
time a significant number of objects with NIR photometry and periods
shorter than 5 day, and the very good sampling of our multi-epoch
$K_\mathrm{s}$ light curves.  In the case of the FO pulsators the
agreement with \cite{groe00} is less satisfactory than for the F-mode
Cepheids. This may be due to the advantage of the deeper magnitude limit achieved
by the VMC survey, which allowed us to reach the fainter FO Cepheids
populating the short-period tail of the $PLK_\mathrm{s}$ relation, along
with the much better sampling of our $K_\mathrm{s}$ light curves.
The \citet[][]{groe00} relations rely in fact on 2MASS and DENIS single
epoch NIR data, thus the Cepheid's mean magnitude is, in
principle, less well  determined increasing, in turn, the r.m.s.  of the
$PLK_\mathrm{s}$ relation. This is a natural consequence of
  studying Cepheids in the NIR using only single epoch data.
\\

Since there are no empirical $PW$ and $PLC$ relations in $K_\mathrm{s}$ 
available in the literature we can only compare our relations with the
theoretical results by \citet{Caputo00}.
% because we lack
%similar empirical relations in the literature.
The slope of our $W(V,K_\mathrm{s})$ relation is in agreement with the
theoretical value, while a significant discrepancy is found in the
case of the $PLC$ relation (see Table~\ref{comparison}).  This
discrepancy could be due, at least in part, to uncertainties affecting
the $(V-K_\mathrm{s})$-temperature transformations that might overestimate the
coefficient of the predicted colour term and, in turn, the linear
regression of the corrected magnitude versus period. This effect is
expected to be mitigated in the Wesenheit approach thanks to the
adoption of  the same colour coefficient (in the empirical and theoretical
relations) set by
Cardelli's law. \\

%Finally, let us take into account the possibility that the various
%PL,PW,PLC derived here do show a sort of discontinuity around 10 days
%as claimed by several authors. 

Finally, we note that no statistically significant evidence is found for the $PL$
discontinuity around 10 day claimed by several
authors. \citet{Ngeow05} performed a statistical study of the
LMC Cepheid sample obtained from the MACHO database and found that the
observed behaviour in the period-magnitude diagrams is best reproduced by
two linear relations, with a break at 10 day. A similar deviation
from linearity is predicted by nonlinear convective pulsation models
\citep[see e.g.][]{Caputo00,Fiorentino02,mmf05,m10}, which also suggest
a quadratic form of the $PL$ relations,  particularly in the optical bands.
However, all these authors predict that the NIR $PL$ and $PW$ relations are
well approximated by linear relations \citep[see also][]{Ngeow08}, as we confirm with the present  
study.

\section{Classical Cepheids in the SEP field}

\begin{figure}
\includegraphics[width=8cm]{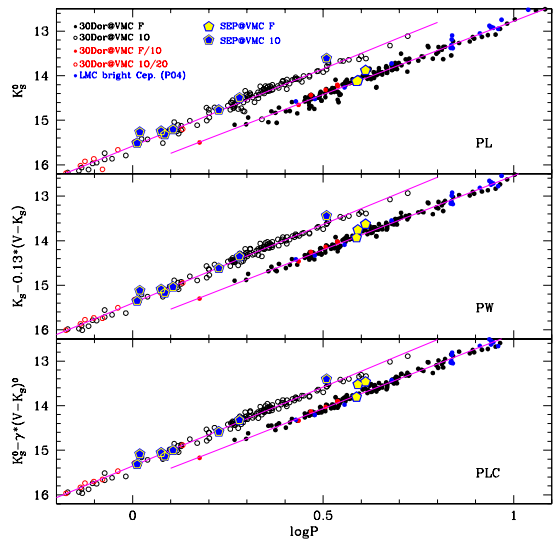}
\caption{Classical Cepheids (blue and yellow pentagons for F and
FO pulsators, respectively) in the SEP field overplotted on the $PL$, $PW$ and $PLC$ 
relationships defined by the 30 Dor Cepheids.}
 %{\bf Nella figura ci vanno scritte l'eq. delle due regressioni lineari)}}
\label{gaiaField}
\end{figure}

The number of SEP  Cepheids is too small  (11 objects, see Table~\ref{tabSep}) 
to define independent $PL$, $PW$, and $PLC$ relations. We did not attempt to re-calculate 
the relations but simply compared  the SEP Cepheids with those obtained  from the 30
Dor variables. This is done in Fig.~\ref{gaiaField}, where for the SEP Cepheids we
have adopted the reddening value $E(B-V)$=0.06 mag from
\citet{Rubele12}\footnote{No  evaluation of the  SEP field reddening is available from 
the \citet{haschke} study.}. 

%When plotted 
 %on the relations defined by the 30 Dor Cepheids (see Fig.~\ref{gaiaFig.ps}) they show some scatter. However, it is not clear whether the scatter is 
%due to them defining different relationships, or to projection effects.
%The results in the $K_\mathrm{s}$ band for the 12 Cepheid stars found by EROS-2 in the SEP field are shown in
%Tab.~\ref{tabSep}. The poor number of objects is justified  by the
%position of this field, far from the bar of the LMC. It is also remarkable the
%predominance of FO pulsators (8) with respect to F
%ones, as well as the absence of pulsators with period longer than 4
%d. This occurrence is probably reflecting a particular SFH of the
%region we are analyzing. \\
%Given the low statistics we did not recalculate the PL, PW and PLC relations in the SEP field. However, it is
%instructive to compare the SEP field Cepheid positions in the PL, PW
%and PLC planes, with the relations obtained  in the 30 Dor field. 

The SEP variables overlap well with the 30 Dor Cepheids in all panels of the figure. 
However, there are
a few exceptions:  stars VMC J055711.13-655116.1, VMC J055638.33-660302.5, 
VMC J055613.37-662234.0 and  VMC J055922.13-665709.6  deviate by more than 
2 $\sigma$ from the $PL$,
$PW$ and $PLC$ relations obtained in the 30 Dor field, and appear to be closer to us by a few kpc. 
Indeed, from the analysis of the CMD data \citet{Rubele12} find the SEP field
to be located on average 0.05 mag closer to us than the 30 Dor region, which is
fully consistent  with the results we find here from the Cepheids.

\begin{table*}
\small
\caption{Literature values for the coefficients of the $PL$, $PW$ and $PLC$ relations, for F and FO Classical
Cepheids.  The Wesenheit function is
defined as:
$W(V,K_\mathrm{s})=K_\mathrm{s}-0.13(V-K_\mathrm{s})$. Note that the photometry of previous
studies was converted to the 2MASS system, for consistency with
our results (see Section 2).}
\label{comparison}
\begin{center}
\begin{tabular}{ccccccccl}
\hline
\noalign{\smallskip}
mode & $\alpha$ & $\sigma_{\alpha}$ & $\beta$ & $\sigma_{\beta}$ & $\gamma$ & $\sigma_{\gamma}$ &r.m.s. & source \\
\noalign{\smallskip}
\hline
\noalign{\smallskip}
\multicolumn{9}{c}{$K_\mathrm{s}^0$=$\alpha$+$\beta$ log$P$} \\
\noalign{\smallskip}
\hline
\noalign{\smallskip}
F   & 16.032 & 0.025 & -3.246& 0.036&  & & 0.168&\citet{groe00}$^{\rm (a)}$  \\ 
F   & 16.051 & 0.05   &  -3.281& 0.040&  & & 0.108&\citet{persson}$^{\rm (b)}$ \\
F   & 15.945 & 0.040   &  -3.19& 0.040&  & & &\citet{Testa07}$^{\rm (b)}$ \\
F   & -2.36 & 0.04   & -3.28   & 0.09  &  &  &  0.21 & \citet{Storm11b} $^{\rm (c)}$\\
F   & -2.65 & 0.01 & -3.23& 0.01&  & & 0.07&\citet{Caputo00}$^{\rm (d)}$  \\   
FO & 15.533& 0.032 & -3.381& 0.076&  & & 0.137&\citet{groe00} $^{\rm (a)}$ \\
FO & 15.62& 0.13 & -3.57& 0.03&  & & 0.14&\citet{Bono02}$^{\rm (a)}$  \\
\noalign{\smallskip}
\hline
\noalign{\smallskip}
\multicolumn{9}{c}{$W(V,K_\mathrm{s})=\alpha+\beta $ log$P$} \\
\noalign{\smallskip}
\hline
\noalign{\smallskip}
F   & -2.92 & 0.09 & -3.21& 0.04&  & & 0.09&\citet{Caputo00}  \\  
\noalign{\smallskip}
\hline
\noalign{\smallskip}
\multicolumn{9}{c}{$K_\mathrm{s}^0=\alpha+\beta$ log$P$$+\gamma (V-K_\mathrm{s})_0$} \\
\noalign{\smallskip}
%\hline
\noalign{\smallskip}
F   & -3.37 & 0.04 & -3.60 & 0.03 & 0.61& 0.03& 0.03  &\citet{Caputo00} \\
%FO & 15.355 & 0.070 & -3.545 & 0.026 & 0.163& 0.014 & 0.070 & \citet{Bono02}\\
\noalign{\smallskip}
\hline
\noalign{\smallskip}
\multicolumn{9}{l}{$^{\rm (a)}$ data in the CIT system: $K_\mathrm{s}$(2MASS)=$K$(CIT)$-0.024$ \citep[][]{carpenter}} \\
\multicolumn{9}{l}{$^{\rm (b)}$ data in the LCO system: $K_\mathrm{s}$(2MASS)$=K_\mathrm{s}$(LCO)$-0.01$ \citep[][]{carpenter}} \\
\multicolumn{9}{l}{$^{\rm (c)}$ data in the SAAO system: $K_\mathrm{s}$(2MASS)$=K$(SAAO)$+0.02(J-K)$(SAAO)$-0.025$ \citep[][]{carpenter}} \\
\multicolumn{9}{l}{$^{\rm (d)}$ models transformed to the Johnson system:
  for Cepheids $K$(Johnson) $\approx K$(SAAO) \citep[][]{bessellbrett88} }\\
\end{tabular}
\end{center}
\end{table*}

\section{Calibration of the $PL$, $PW$ and $PLC$ zero points}% and the distance of LMC}

Our main goal is to use the Classical Cepheids observed by the VMC (along with the RR Lyrae stars, M12)  to trace the
3D geometry of the Magellanic system. However,  this analysis would be premature with the few VMC tiles 
observed so far.  Still,  we can use the $PL$, $PW$ and $PLC$ relations derived in the previous section
to estimate an absolute distance to the LMC,  but first need to  calibrate their zero points.
In the following we shall use mainly the $PW$ and $PLC$ relations, as they appear to  be less
dispersed than the $PL$ relation, and derive zero point
estimates, adopting for the first time the $V, (V-K_\mathrm{s})$ combination,  and 
two ``direct techniques'':  the trigonometric parallaxes of Galactic Cepheids,  and 
the Baade-Wesselink method  directly applied  to LMC Cepheids. 
% with the data
%available at the moment, this analysis would be premature. Nevertheless,
%the PL, PW and PLC relations derived in the previous section can be
%used to estimate the absolute distance of the LMC. To this aim we
%mainly rely on the PW and PLC relations, that showed to be less
%dispersed than the PL,  and provide zero point
%estimates, for the first time adopting the $V, (V-K)$ combination, using
%two ``direct techniques'': 
%Whatever the method to use, it is useful to
%calibrate the zero points of the quoted relation, both because it has
%never been done in the literature and also to use them to find the
%absolute distance to the LMC. To this aim we have different choices.
 % We decided to rely only to ``direct'' measures of the zero
%points, i.e. 
%the trigonometric parallaxes (of Galactic Cepheids) and 
%the Baade-Wesselink method directly applied to LMC pulsators.    
%In the two following subsections we illustrate the results obtained
%from the application of these two methods.

\begin{table*}
\scriptsize
\caption{Galactic Cepheids with know parallax, used to calibrate the $PL$, $PW$ and
 $PLC$ relations.  The column named ``LK'' gives the Lutz-Kelker
 corrections applied in this work. An ``FO'' in the notes means that the star is a suspect
 first overtone, according to the compilation by \citet{Fernie95}.
``O'', ``B'' and ``V'' mean that the star is a binary with known orbital elements,
 spectroscopic and  visual, respectively; a ``:'' means that 
 confirmation is needed. After \citet{Szabados03}.}
\label{galactic}
\begin{center}
\begin{tabular}{lcccccccc}
\hline
\noalign{\smallskip}
ID & $\pi$  & $\sigma_\pi$ & log$P$ & $\langle V \rangle$ & $\langle K \rangle$ & $E(B-V)$ & LK & note\\
 & \arcsec  & \arcsec  & d & mag & mag & mag & mag  & \\
\noalign{\smallskip}
\hline
\noalign{\smallskip}
SU Cas        &    2.57    &     0.33   &     0.440    &    5.9700 &4.1062   &     0.2590   &    -0.1649  & FO, O \\  
$\beta$ Dor   &    3.26    &     0.14   &     0.993    &    3.7570    &    1.9430   &     0.0520   &    -0.0184 & \\  
RT Aur        &    2.40    &     0.19   &     0.572    &    5.4480 &    3.8962   &     0.0590   &    -0.0627  & B:\\  
$\zeta$ Gem   &    2.74    &     0.12   &     1.006    &    3.9150&2.1145   &     0.0140   &    -0.0192 & V\\  
$\ell$ Car    &    2.03    &     0.16   &     1.551    &    3.6980    &    1.0788   &     0.1470   &    -0.0621  &\\  
BG Cru        &    2.23    &     0.30   &     0.678    &    5.4590 &    3.8704   &     0.1320   &    -0.1810 &FO, B \\  
X Sgr         &    3.17    &     0.14   &     0.846    &    4.5640&    2.5033   &     0.2370   &    -0.0195  &O \\  
W Sgr         &    2.30    &     0.19   &     0.880    &    4.6700    &    2.8101   &     0.1080   &    -0.0682 & O\\  
Y  Sgr        &    2.13    &     0.29   &     0.761    &    5.7450    &    3.5695   &     0.1910   &    -0.1854  &B\\  
FF Aql        &    2.64    &     0.16   &     0.650    &    5.3730    &    3.4706   &     0.1960   &    -0.0367  &O\\  
T Vul         &    2.06    &     0.22   &     0.647    &    5.7530    &    4.1814   &     0.0640   &    -0.1141  &B\\  
DT Cyg        &    2.19    &     0.33   &     0.550    &    5.7750    &    4.4109   &     0.0420   &    -0.2271 &FO \\  
$\delta$ Cep  &    3.71    &     0.12   &     0.730    &    3.9530    &    2.3037   &     0.0750   &    -0.0105 &V \\  
\noalign{\smallskip}
\hline
\noalign{\smallskip}
\end{tabular}
\end{center}
\end{table*}

\subsection{Zero points from the parallax of Galactic Cepheids}

Accurate parallaxes for Galactic Cepheids are available for fewer than
20 objects, from observations with the Hipparcos satellite
\citep{vanLee07}, and the HST
\citep[][]{Benedict07}. To use these data for our
purpose, we had to: i) derive the $K$-band $PL$ relation, and, for the first time, the 
$V, (V-K_\mathrm{s})$- $PW$ and $PLC$ relations for these Cepheids; ii) apply them
to our sample taking into account that possible differences may exist
in the slopes and zero points, due to metallicity effects.

\begin{figure}
\includegraphics[width=8cm]{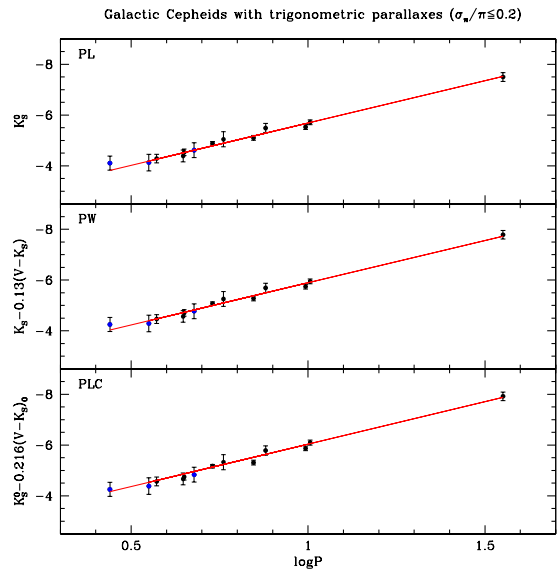}
\caption{$PL$,$PW$ and $PLC$  relations for Galactic Cepheids with accurate 
trigonometric parallax. We have highlighted in blue the suspected
FO-mode pulsators (see Table~\ref{galactic}).}
\label{trigPar}
\end{figure}

From the \citet{vanLee07} and \citet{Benedict07} samples we  only retained
stars with the most accurate parallax ($\delta\pi/\pi\leq0.2$). For 10
stars in common between the two sets we computed weighted averages of
the parallaxes.  Photometric data \citep[including $K$-band
photometry, transformed to the 2MASS $K_\mathrm{s}$
 system according to][]{carpenter}
and individual reddening values for these stars were taken from
\citet{Fouque07}, while for the Lutz-Kelker corrections we followed
\citet{Benedict07}.  We then computed the $PL$, $PW$ and $PLC$
relations, by excluding from the fit the most deviating stars:
$\alpha$ UMi and S Mus.  This left us with a total number of 13
Galactic Cepheids. Their list is provided in Table~\ref{galactic}.
Inclusion in the fit of the three suspected FO stars (see
Table~\ref{galactic}) did not change the results significantly, hence
we kept these stars to increase the statistics. We also kept all
binary objects (see Table~\ref{galactic}). 
Excluding them would considerably reduce the statistical
significance of our results. This limitation, due to the paucity of
trigonometric parallaxes for Cepheids, will be directly addressed when the
astrometric satellite Gaia is launched and goes into operation in 2013.
%
%The 13 Cepheids listed in Tab.~\ref{galactic} were then used to calculate the $PL$, $PW$ and $PLC$
%relations. In particular, 
We first calculated the regressions leaving all parameters free to
vary. The colour term in the $PLC$ relation turned out to be 
insignificant, thus the $PL$ and $PLC$ relationships are identical and equal to:
$K^0_S=-2.44\pm0.12-(3.20\pm0.14)$log$P$.  Similarly, for the Wesenheit
function we have: $W(V,K_\mathrm{s})=-2.61\pm0.12-(3.28\pm0.13)$log$P$.
Although rather uncertain, the slopes of these relations are in good
agreement with our results from the 30 Dor Cepheids, within the
errors.  We thus adopted our slopes from the 30 Dor Cepheids for the
$PL$ and $PW$ relations to derive the following weighted-average zero
points of the parallax-based relations:

\begin{equation}
K^0_\mathrm{s}{\rm (F)}=-2.40\pm0.05-(3.295\pm0.018){\rm log } P
\label{eqPW} 
\end{equation}

\begin{equation}
W(V,K_\mathrm{s}){\rm (F)}=-2.57\pm0.05-(3.325\pm0.014) {\rm log } P 
\label{eqPW} 
\end{equation}
\noindent
where the error on the zero point is the standard deviation of the
mean. Similarly,  by adopting our values of $0.216\pm0.014$ and
$-3.346\pm0.013$ for the period and colour coefficients of the $PLC$ relation, we obtained:

\begin{eqnarray}
K^0_\mathrm{s}&=&-2.69\pm0.05-(3.346\pm0.013){\rm log } P \nonumber \\
&&+(0.216\pm0.014)(V-K_\mathrm{s})_0 
\label{eqPLC} 
\end{eqnarray}
\noindent
Here,  the errors on the zero point estimates include the
contribution of the 
systematic uncertainty due to the adoption of the same slope for Galactic and 
LMC Cepheids according to the model predictions by \citet{Caputo00}. 
%A visualization of the previous equations is displayed in Fig.~\ref{trigPar}.
The $PL$, $PW$ and $PLC$ relations obtained with this procedure are
shown in Fig.~\ref{trigPar}.  By comparing Eqs.~\ref{eqPW} and
~\ref{eqPLC} with the results in Table~\ref{tab1}, we obtain distance
moduli of the LMC, based on the F-mode Cepheids, of:
$(m-M)^{\rm TRIG}_0(PW)=18.44\pm0.05$ mag, and
$(m-M)^{\rm TRIG}_0(PLC)=18.43\pm0.05$ mag, respectively.
%
%can directly obtain two estimates of the distance modulus of the
%LMC, namely $(m-M)^{TRIG}_0(PW)=18.44\pm0.05$ and
%$(m-M)^{TRIG}_0(PLC)=18.43\pm0.05$, 

%    error on metallicity 0.04 mag summed in quadrature with the others

\subsection{Zero points from the Baade-Wesselink method}

%\\

\subsubsection{Infrared Surface Brightness (IRSB) formulation}

Recently, \citet{Storm11a,Storm11b} published individual
distances to Galactic and LMC Cepheids (36 objects in the LMC), based
on application of the IRSB %InfraRed Surface Brightness
modification of the Baade-Wesselink technique \citep[see e.g.][and
references therein]{Gieren07} to a sample of Galactic and LMC Cepheids
and a new evaluation of the projection factor (p-factor) that
transforms the observed radial velocity into pulsation velocity.
%, to obtain the individual distances to the 36 objects. We then took their absolute
We have used their  absolute magnitudes in 
$K$\footnote{Note that these $K$ magnitudes were transformed to the
  2MASS $K_\mathrm{s}$ system and referred to the centre of the LMC using the
  corrections by \citet{Storm11b}.} and $V$ to calibrate the zero points of the $PL$, $PW$ and $PLC$ relations. 
%
%$K$\footnote{Note that these $K$ magnitudes were transformed to the
%  2MASS $K_\mathrm{s}$ system and referred to the centre of LMC using the
%  corrections by \citet{Storm11b}.} and $V$
%magnitudes to calibrate the zero points of our PW and PLC relations. 
As done with the Galactic Cepheids, we adopted our slopes for the
relations, and determined their zero points.  The assumption is well
justified in this case as both our and \citet{Storm11b}'s Cepheids
belong to the LMC, hence we do not expect any metallicity effect.
%since both our and \citet{Storm11b} samples belong to the LMC. 
As a result we obtain:

\begin{equation}
K^0_\mathrm{s}{\rm (F)}=-2.40\pm0.07-(3.295\pm0.018){\rm log } P
\label{eqPW1} 
\end{equation}

\begin{equation}
W(V,K_\mathrm{s}){\rm (F)}=-2.60\pm0.07-(3.325\pm0.014){\rm log } P
\label{eqPW1} 
\end{equation}

%\begin{eqnarray}
%K^0_\mathrm{s}(F)&=&-2.72\pm0.07-(3.346\pm0.013){\rm log }$P$ \nonumber \\
%&&+(0.216\pm0.014)(V-K_\mathrm{s})_0
%\label{eqPLC1}
%\end{eqnarray}

\begin{eqnarray}
%\begin{split}
\noindent
   K^0_\mathrm{s}(F)&=&-2.72\pm0.07-(3.346\pm0.013){\rm log } P \nonumber \\
&&+(0.216\pm0.014)(V-K_\mathrm{s})_0  
 %  \end{split}
\end{eqnarray}
\noindent
where the errors on the zero points include the dispersion of the
measurements, as well as an estimate of the systematic error due to
the uncertainty on the p-factor \citep[see e.g.][and references
therein]{Storm11a}.  The new zero points are in excellent agreement
with those derived from the Galactic Cepheids and confirm
\citet{Storm11a}'s findings.
%This is not surprising because it was
%already found by \citet{Storm11a}. 
%As for the distance, we find
These relations lead to distance moduli for the LMC of:
$(m-M)^{\rm IRSB}_0(PW)=18.46\pm0.07$ mag and
$(m-M)^{\rm IRSB}_0(PLC)=18.46\pm0.07$ mag, respectively, in very good
agreement with \citet{Storm11b}.
%.This result differs by only
%0.01 mag from that derived by \citet{Storm11b}. This is not surprising as
% we used the same calibrators. 
%accounted for 0.06 mag due to p-factor
\\
%\subsection{Results from the CORS Baade-Wesselink method}

\subsubsection{CORS Baade-Wesselink formulation}

A different implementation of the Baade-Wesselink technique, the
so-called CORS method \citep[see e.g.][]{Cors81,Ripepi97,Molinaro11}, 
was applied by \citet{Molinaro12} to 9 Cepheids (7 F- and 2 FO-mode)
belonging to NGC1866, a populous widely studied young cluster located
about 4.1\degr North--West of the LMC bar.
%in the LMC
%, deriving for the cluster 
%
%on a sample (9 stars, 7 F- and 2 FO-mode) of Cepheids belonging to the blue populous  
%cluster NGC1866, a widely studied LMC object located about 4.1 \degr
%North-West with respect to the galaxy bar. 
These authors find $(m-M)_0$(NGC1866)=$18.51\pm0.03$ mag. This value can be
used to calibrate the $PL$, $PW$ and $PLC$ relations following the
same procedure described in the previous sections, and with the
advantage that we can now calibrate also the relation for FO pulsators
(although only based on two stars). In this case we obtain:
%
% our PW and PLC exactly as done in the previous two subsections. We
% note that in this case we can also calibrate the relations
%for FO-mode pulsators (even if the zero points rely only on two
%stars). Finally we obtain:

%\begin{eqnarray}
%K^0_S(F) & = & -2.44\pm0.07-(3.295\pm0.018){\rm log } $P$ \\
%K^0_S{\rm (FO)} & = & -2.94\pm0.07-(3.471\pm0.035){\rm log } $P$
%\end{eqnarray}

\begin{equation}
K^0_\mathrm{s}{\rm (F)}  =  -2.44\pm0.07-(3.295\pm0.018){\rm log } P 
\end{equation}

\begin{equation}
K^0_\mathrm{s}{\rm (FO)}  =  -2.94\pm0.07-(3.471\pm0.035){\rm log } P
\end{equation}

\begin{equation} 
W(V,K_\mathrm{s}){\rm (F)} = -2.62\pm0.07-(3.325\pm0.014){\rm log } P 
\end{equation}

\begin{equation} 
W(V,K_\mathrm{s}){\rm (FO)} = -3.10\pm0.07-(3.530\pm0.025){\rm log } P
\end{equation}

\begin{eqnarray}
{K^0_\mathrm{s}}{\rm (F)}&=&-2.74\pm0.07-(3.346\pm0.013){\rm log } P \nonumber \\
&&+(0.216\pm0.014)(V-K_\mathrm{s})_0 
\end{eqnarray}

\begin{eqnarray}
{K^0_\mathrm{s}}{\rm (FO)}&=&-3.15\pm0.07-(3.545\pm0.026){\rm log } P \nonumber\\
&&+(0.163\pm0.014)(V-K_\mathrm{s})_0
\end{eqnarray}

\noindent
Here, as in the previous section, the errors in the zero points
include both the dispersion of the measures, and the uncertainty in
the p-factor. 

Averaging the results for F and FO Cepheids, we obtain distance moduli
of: $(m-M)^{\rm CORS}_0(PW)=18.49\pm0.07$ mag and
$(m-M)^{\rm CORS}_0(PLC)=18.49\pm0.07$ mag, respectively.  The
marginally significant $-$0.02 mag difference with \citet{Molinaro12}
is likely due to the relative distance between NGC1866 and the LMC
centre.  Indeed, a direct comparison of the NGC1866 Cepheids with our
$PL$ relation reveals a difference on the order of $-0.02 \pm$0.02 mag. Such
an uncertainty, when added in quadrature, has negligible effects
compared to other sources of error.

\section{Discussion and estimate of the distance to the LMC}

In the previous sections we have calibrated the zero points of our
$PL$, $PW$ and $PLC$ relations using different methods, and derived
in turn different estimates of the distance to the LMC.
% , whose values, in any case, strictly depend on the adopted
% calibrators.
By combining these results, we can derive our best estimates for the
zero points of the $PL$, $PW$ and $PLC$ relations, as well as for the
distance to the LMC.  In particular, a weighted mean of the zero
points for F-mode pulsators leads to the following final $PL$, $PW$
and $PLC$ relations:

\begin{equation}
K^0_\mathrm{s}{\rm (F)}  =  -2.41\pm0.03-(3.295\pm0.018){\rm log } P 
\label{plFin}
\end{equation}

\begin{equation} 
W(V,K_\mathrm{s}){\rm (F)} = -2.59\pm0.03-(3.325\pm0.014){\rm log } P 
\label{pwFin}
\end{equation}

\begin{eqnarray}
{K^0_\mathrm{s}}{\rm (F)}&=&-2.71\pm0.03-(3.346\pm0.013){\rm log } P \nonumber \\
&&+(0.216\pm0.014)(V-K_\mathrm{s})_0 
\label{plcFin}
\end{eqnarray}

\noindent
As for the distance to the LMC, a similar weighted mean leads to the
final value of $(m-M)_0=18.46\pm0.03$ mag, in excellent agreement with
several independent literature results \citep[see, e.g.,][and references therein]{w12}. In particular, \citet{w12} in his
recent review of current LMC distance estimates based on Cepheids, red
variables, RR Lyrae, Red Clump stars and Eclipsing Binaries, showed
that most of these indicators agree on a mean value of the distance
modulus for the LMC of 18.48$\pm$ 0.05 mag.

%Our new final distance modulus for the LMC is slightly shorter than the value assumed 
%by the HST Key Project (18.50 mag), thus implying  LMC calibrated extragalactic
%distances shorter by about 2 \%, a small but still non negligible
%correction,  in the era of  the claimed few \% Hubble constant precision
%\citep[see e.g.][]{Rie11}. 

\section{Summary and Conclusions}

We have presented first results for Classical Cepheids observed by the
VMC survey in two fields of the LMC, centred on the SEP and the 30
Dor regions, respectively. The identification of the variables and their 
optical magnitudes were derived from the EROS-2 and OGLE-III
catalogues.  Our Cepheid $K_\mathrm{s}$ light curves are very well
sampled, with at least 12 epochs, and very precise, with typical
errors of 0.01 mag, or better, for individual phase points. Our
observing strategy allowed us to measure for the first time the
$K_\mathrm{s}$ magnitude of the faintest Cepheids in the LMC, which
are mostly FO pulsators, thus enabling us to obtain $PL$, $PW$ and
$PLC$ relations defined over the whole period range spanned by the LMC
FO Cepheids.  
%Saturation limits our $K_\mathrm{s}$ measurements of the
%Fundamental mode (F) Cepheids to periods shorter than 15-20 days.
Since the longest period  Cepheid in our dataset is a 23 day variable,
%
%included in our sample is around 23 days,}
we have complemented our sample with data from
\citet{persson} for bright F-mode Cepheids. On this basis we have
built a $PL$ relation in the $K_\mathrm{s}$ band that for the first time
includes also short period (i.e. low luminosity) F pulsators. We have
also calculated the first empirical $PL$, $PW$ and $PLC$ relations
using the $(V-K_\mathrm{s})$ colour. The latter two have very small
dispersion ($\leq0.07$ mag) and will be particularly suited for the 3D
study of the Magellanic system, a main goal of the VMC project, that
we cannot yet achieve with only two VMC fields completed so far.
%
%We have turned instead 
%face in this paper. However, given that we have only two VMC fields completed, 
%this problem was not faced in this paper. Instead, 
We have used ``direct'' distance measures to both Galactic and LMC
Cepheids to calibrate the zero points of the $PL$, $PW$ and $PLC$ relations
derived in this paper. This procedure led to
Eqs. ~\ref{plFin}--\ref{plcFin} that we have used to
estimate an absolute distance to the LMC of $(m-M)_0=18.46\pm0.03$ mag, in
excellent agreement with the latest determinations of the LMC distance,
based on Classical Cepheids
and other independent distance indicators. \\

The new final distance modulus for the LMC derived in the present
study is slightly shorter than the value assumed by the HST Key
Project \citep[18.50 mag, according to][]{f01}, thus implying LMC-calibrated
extragalactic distances shorter by about 2 \%, a small but still non
negligible correction, in the era of the claimed Hubble constant
within a few per cent uncertainty \citep[see e.g.][]{Rie11}.

%The smaller deviation from
%the currently assumed value for the LMC distance scale  in the
%application of extragalactic Cepheid distances to infer the Hubble
%constant $H_0$, would imply  Cepheid distances ($H_0$) shorter
%(higher) by 2\% than the values inferred by the HST Key Project \citep[][]{f01}.

\section*{Acknowledgments}

It is a pleasure to thank our referee B. Madore for his prompt and
helpful report.  V.R. warmly thanks Roberto Molinaro for providing the
program for the spline interpolation of the light curves.
M.I. Moretti thanks the Royal Astronomical Society for financial
support during her two-month stay at the  University of Hertfordshire. 
Financial support for this work was provided by PRIN-INAF 2008
(P.I. Marcella Marconi) and by COFIS ASI- INAF I/016/07/0.
RdG acknowledges partial research support from the National Natural
Science Fundation of China (grant 11073001). 
We thank the UK's VISTA Data Flow System comprising the VISTA pipeline
at the Cambridge Astronomy Survey Unit (CASU) and the VISTA Science
Archive at Wide Field Astronomy Unit (Edinburgh) (WFAU) provided
calibrated data products, and is supported by STFC.

%\appendix

%\section[]{}

%\label{lastpage}


\begin{thebibliography}{99}

\bibitem[\protect\citeauthoryear{Bekki \& Chiba}{2007}]{bekki07}
Bekki K., Chiba M., 2007, MNRAS, 381, L16

\bibitem[\protect\citeauthoryear{Benedict et al.}{2007}]{Benedict07}   
Benedict G. F., McArthur B. E., Feast M. W., Barnes 
T. G., Harrison T. E., Patterson  R. J., Menzies J. W., 
Bean  J. L., Freedman  W. L., 2007, AJ, 133, 1810

\bibitem[\protect\citeauthoryear{Bessell \& Brett}{1988}]{bessellbrett88}   
Bessell, M. S., Brett, J. M., 1988, PASP, 100, 1134

\bibitem[\protect\citeauthoryear{Bono et al.}{1999}]{bccm99}   
Bono G., Caputo F., Castellani V., Marconi M., 1999, ApJ, 512, 711

\bibitem[\protect\citeauthoryear{Bono et al.}{2002}]{Bono02}   
Bono G., Groenewegen M. A. T., Marconi M., Caputo F.,  2002, ApJ,
574, L33

\bibitem[\protect\citeauthoryear{Bono et al.}{2008}]{b08}   
Bono G., Caputo F., Marconi M., Musella I., 2008, ApJ, 684, 102

\bibitem[\protect\citeauthoryear{Bono et al.}{2010}]{b10}   
Bono G., Caputo F., Marconi M., Musella I., 2010, ApJ, 715, 277

\bibitem[\protect\citeauthoryear{Borissova et al.}{2009}]{borissova}  
Borissova J., Rejkuba M., Minniti D., Catelan M., Ivanov, V. D., 2009, A\&A, 502, 505 

\bibitem[\protect\citeauthoryear{Brocato et al.}{2004}]{bro04}  
Brocato E., Caputo F., Castellani V., Marconi M., Musella I., 2004, AJ, 128, 1597


\bibitem[\protect\citeauthoryear{Caccin et al.}{1981}]{Cors81} 
Caccin R., Onnembo A., Russo G., Sollazzo C., 1981, A\&A, 97, 104

\bibitem[\protect\citeauthoryear{Caputo, Marconi \& Musella}{2000}]{Caputo00}
Caputo F., Marconi M., Musella I., 2000, A\&A, 354, 610

\bibitem[\protect\citeauthoryear{Cardelli, Clayton \& Mathis}{1989}]{cardelli89}
Cardelli J. A., Clayton  G. C., Mathis J. S., 1989, ApJ, 345, 245

\bibitem[\protect\citeauthoryear{Carpenter}{2001}]{carpenter} 
Carpenter J.M., 2001, AJ, 121, 2851 

\bibitem[\protect\citeauthoryear{Cioni et al.}{2011}]{Cioni11}
Cioni M.-R. L., Clementini G., Girardi L., et al., 2011,
A\&A, 527, 116 (Paper I)

\bibitem[\protect\citeauthoryear{Coppola et al.}{2011}]{coppola11}
Coppola G., Dall'Ora M., Ripepi V., et al., 2011, MNRAS, 416,1056


\bibitem[\protect\citeauthoryear{Cross et al.}{2009}]{Cross09}
Cross N. J. G., Collins R. S., Hambly N. C., Blake R. P., Read M. A.,
Sutorius E. T. W. and Mann R.G., Williams P. M., 2009, MNRAS, 339, 1730 

\bibitem[\protect\citeauthoryear{Cross et al.}{2012}]{Cross12}
Cross N.J.G., et al., 2012, MNRAS, submitted 


\bibitem[\protect\citeauthoryear{Dall'Ora et al.}{2004}]{dallora04}
Dall'Ora M., Storm J., Bono G., et al., 2004, ApJ, 610, 269

\bibitem[\protect\citeauthoryear{Dalton et~al.}{2006}]{Dalton_etal06}
{Dalton} G. B., {Caldwell} M., {Ward} A.K., {et~al.} 2006, in Society of
 Photo-Optical Instrumentation Engineers (SPIE) Conference Series, Vol. 6269,
 Society of Photo-Optical Instrumentation Engineers (SPIE) Conference Series

\bibitem[\protect\citeauthoryear{Di Criscienzo et al.}{2006}]{dc06}
Di Criscienzo M., Caputo F., Marconi M., Musella, I., 2006, MNRAS,
365, 1357

\bibitem[\protect\citeauthoryear{Emerson et al.}{2004}]{Emerson_etal04} 
Emerson J. P., Irwin M. J., Lewis J., et al., 2004,SPIE, 5493, 401, 41

\bibitem[\protect\citeauthoryear{Emerson, McPherson \& Sutherland}{2006}]{Emerson_etal06}
{Emerson} J. P., {McPherson} A., {Sutherland} W., 2006, The
Messenger, 126, 41

\bibitem[\protect\citeauthoryear{Fernie et al.}{1995}]{Fernie95}
 Fernie J. D., Beattie B., Evans N. R., Seager S., 1995, IBVS, No. 4148

\bibitem[\protect\citeauthoryear{Fiorentino et al.}{2002}]{Fiorentino02}
 Fiorentino G., Caputo F., Marconi M., Musella I., 2002, ApJ, 576, 402



\bibitem[\protect\citeauthoryear{Freedman et al.}{2001}]{f01}
 Freedman, W. L. et al., 2001, ApJ, 553, 47

\bibitem[\protect\citeauthoryear{Freedman \& Madore}{2011}]{fm11}
 Freedman W. L., Madore B. F., 2011, ApJ, 734, 46

\bibitem[\protect\citeauthoryear{Fouqu\'e et al.}{2007}]{Fouque07}
 Fouqu\'e P., Arriagada P., Storm J., Barnes T. G., Nardetto N.,
 M\'erand A., Kervella 
P., Gieren W., Bersier D., Benedict G. F., McArthur B. E., 2007, A\&A, 476, 73

\bibitem[\protect\citeauthoryear{Gieren, Fouqu\'e \& Gomez}{2007}]{Gieren07}
Gieren W. P., Fouqu\'e P., Gomez, M. I., 1997, ApJ, 488, 74


\bibitem[\protect\citeauthoryear{Groenewegen}{2000}]{groe00}
Groenewegen M. A. T., 2000, A\&A, 363, 901

%\bibitem[\protect\citeauthoryear{Hambly et al.}{2008}]{ham08}
%Hambly N. C., Collins R. S., Cross N. J. G., et al., 2008, MNRAS, 384, 637

\bibitem[\protect\citeauthoryear{Harris \& Zaritsky}{2004}]{HZ04}
Harris J., Zaritsky D., 2004, AJ, 127, 1531

\bibitem[\protect\citeauthoryear{Harris \& Zaritsky}{2009}]{HZ09}
Harris J., Zaritsky, D., 2009, AJ, 138, 1243

\bibitem[\protect\citeauthoryear{Haschke, Grebel \& Duffau}{2011}]{haschke}
Haschke R., Grebel E. K., Duffau, S., 2011, AJ, 141, 158 

\bibitem[\protect\citeauthoryear{Irwin et al.}{2004}]{Irwin_etal04}
Irwin M. J., Lewis J., Hodgkin S., et al., 2004, SPIE, 5493, 411

\bibitem[\protect\citeauthoryear{Lindegren \& Perryman}{1996}]{Lin96} 
Lindegren L., Perryman, M. A. C., 1996, A\&AS, 116, 579

\bibitem[\protect\citeauthoryear{Lindegren}{2010}]{Lin10} Lindegren L., 2010, IAU Symposium, 261, 296

%\bibitem[\protect\citeauthoryear{Longmore, Fernley, \& Jameson}{1986}]{long86}
%Longmore A. J., Fernley J. A., Jameson R. F., 1986, MNRAS, 220, 279

\bibitem[\protect\citeauthoryear{Madore}{1982}]{m82}
Madore B. F., 1982, ApJ, 253, 575

\bibitem[\protect\citeauthoryear{Madore \& Freedman}{1991}]{mf91}
Madore B. F. , Freedman W., 1991, PASP, 103, 933

\bibitem[\protect\citeauthoryear{Marconi}{2009}]{m09}
Marconi M., 2009, MmSAI, 80, 141

\bibitem[\protect\citeauthoryear{Marconi, Musella,  \& Fiorentino}{2005}]{mmf05}
Marconi M., Musella I., Fiorentino G., 2005, ApJ, 632, 590

\bibitem[\protect\citeauthoryear{Marconi et al.}{2010}]{m10} 
Marconi M.,  Musella I., Fiorentino G.,  et al., 2010, ApJ, 713, 615

\bibitem[\protect\citeauthoryear{Marquette et al.}{2009}]{marque09}
Marquette J.-B., Beaulieu J. P., Buchler J. R., et al., 2009, A\&A, 495, 249

\bibitem[\protect\citeauthoryear{Molinaro et al.}{2011}]{Molinaro11}
Molinaro R., Ripepi V., Marconi M., Bono G., Lub J., Pedicelli
S., Pel J. W., 2011, MNRAS, 413, 942

\bibitem[\protect\citeauthoryear{Molinaro et al.}{2012}]{Molinaro12}
Molinaro R., Ripepi V., Marconi M., Musella I., Brocato E.,
Mucciarelli A., Stetson P. B., Storm J., Walker A. R., 2012, ApJ, 748, 69

\bibitem[\protect\citeauthoryear{Moretti et al.}{2012}]{Moretti12}
Moretti M. I., Clementini G., Ripepi V., et al., 2012, MNRAS, to be
submitted  (M12)

\bibitem[\protect\citeauthoryear{Muller et al.}{2004}]{muller04}
Muller E., Stanimirovi\'c S., Rosolowsky E., Staveley-Smith L., 2004, ApJ, 616, 845

\bibitem[\protect\citeauthoryear{Neilson \& Langer}{2012}]{nl12} 
Neilson H. R., Langer N., 2012, A\&A, 537, 26

\bibitem[\protect\citeauthoryear{Ngeow \& Kanbur}{2005}]{nk05}
Ngeow C.-C., Kanbur S. M., 2005, MNRAS, 360, 1033

\bibitem[\protect\citeauthoryear{Ngeow et al.}{2005}]{Ngeow05}
Ngeow C.-C., Kanbur S. M., Nikolae, S., Buonaccorsi J., Cook K. H.,
Welch D. L., 2005, MNRAS, 363, 831

\bibitem[\protect\citeauthoryear{Ngeow, Kanbur \& Nanthakumar}{2008}]{Ngeow08} 
Ngeow C.-C., Kanbur S. M., Nanthakumar A., 2008 A\&A, 477, 621

\bibitem[\protect\citeauthoryear{Ngeow}{2012}]{n12} 
Ngeow, C.-C., 2012, ApJ, 747, 50

\bibitem[\protect\citeauthoryear{Persson et al.}{2004}]{persson}
Persson S. E., Madore B. F., Krzemi\'nski W., et al., 2004, AJ, 128, 2239 

\bibitem[\protect\citeauthoryear{Putman et al.}{1998}]{putman98}
Putman M. E., et al., 1998, Nature, 394, 752

\bibitem[\protect\citeauthoryear{Riess et al.}{2011}]{Rie11} 
Riess A. et al., 2011, ApJ, 730, 119

\bibitem[\protect\citeauthoryear{Ripepi et al.}{1997}]{Ripepi97} 
Ripepi V., Barone F., Milano F., Russo G., 1997, A\&A, 318, 797

\bibitem[\protect\citeauthoryear{Ripepi et al.}{2012}]{Ripepi12}
Ripepi V., Moretti M. I., Clementini G., Marconi M., Cioni 
M.-R. L., Marquette J. B., Tisserand P., 2012, Ap\&SS, in press; also
arXiv:1202.5863

\bibitem[\protect\citeauthoryear{Romaniello et al.}{2005}]{r05}
Romaniello M., Primas F., Mottini M., Groenewegen M. A. T., Bono G.,
Fran\c{c}ois P., 2005, A\&A, 429, 37 

\bibitem[\protect\citeauthoryear{Romaniello et al.}{2008}]{r08}
Romaniello M., Primas F., Mottini M., Groenewegen M. A. T., Bono G.,
Fran\c{c}ois P., 2008, A\&A, 488, 731  

\bibitem[\protect\citeauthoryear{Rubele et al.}{2012}]{Rubele12} 
Rubele S., Kerber L., Girardi L., et al., 2012, A\&A,  537, 106 (Paper IV)

\bibitem[\protect\citeauthoryear{Saha et al.}{2001}]{s01} 
Saha A., Sandage A., Tammann G. A., Dolphin A. E., Christensen
J., Panagia N., Macchetto F. D., 2001, ApJ, 562, 314

\bibitem[\protect\citeauthoryear{Sandage \& Tammann}{1968}]{st68} 
Sandage A., Tammann G., 1968, ApJ 151, 531

\bibitem[\protect\citeauthoryear{Sandage,  Tammann \& Reindl}{2009}]{str09} 
Sandage A., Tammann G. A., Reindl B., 2009, A\&A, 493, 471S

\bibitem[\protect\citeauthoryear{Soszy\'{n}ski et al.}{2008}]{sos08}
Soszy\'{n}ski I., Poleski R., Udalski A., et al., 2008, Acta Astron., 58, 163

\bibitem[\protect\citeauthoryear{Soszy\'{n}ski et al.}{2009}]{sos09}
Soszy\'{n}ski I, Udalski A., Szyma\'{n}ski M. K., al., 2009, Acta Astron., 59, 1

\bibitem[\protect\citeauthoryear{Stanimirovi\'c et al.}{2004}]{stanim04} 
Stanimirovi\'c S., Staveley-Smith L., Jones P. A.,  2004, ApJ, 604, 176

\bibitem[\protect\citeauthoryear{Storm et al.}{2011a}]{Storm11a}
Storm J., Gieren W., Fouqu\'e P., Barnes T. G., Pietrzy\'{n}ski G.,
Nardetto N., Weber M., Granzer T., Strassmeier K. G., 
2011, A\&A, 534, A94

\bibitem[\protect\citeauthoryear{Storm et al.}{2011b}]{Storm11b}
Storm J., Gieren W., Fouqu\'e P., Barnes T. G., Soszy\'{n}ski I.,
Pietrzy\'{n}ski G., Nardetto N., Queloz D., 2011, A\&A, 534, A95

\bibitem[\protect\citeauthoryear{Szabados}{2003}]{Szabados03}
Szabados L., 2003, IBVS, No. 5394

\bibitem[\protect\citeauthoryear{Szewczyk et al.}{2008}]{szewczyk}  
Szewczyk O., Pietrzy{\'n}ski G., Gieren W., et al., 2008, AJ, 136, 272 

\bibitem[\protect\citeauthoryear{Testa et al.}{2007}]{Testa07} 
Testa V., Marconi M., Musella I., Ripepi V., Dall'Ora M.,
Ferraro F. R., Mucciarelli A., Mateo M., C\^{o}t\'e P., 2007,  A\&A,
462, 599

\bibitem[\protect\citeauthoryear{Tisserand et al.}{2007}]{tisserand07}   
Tisserand P., Le Guillou L., Afonso C., et al., 2007, A\&A, 469, 387 

\bibitem[\protect\citeauthoryear{van Leeuwen et al.}{2007}]{vanLee07}  
van Leeuwen F., Feast M. W., Whitelock P. A., Laney C. D., 2007, MNRAS,
379, 723

\bibitem[\protect\citeauthoryear{van der Marel \& Cioni}{2001}]{vandermarel01}    
van der Marel R. P., Cioni M.-R. L.,  2001, AJ, 122, 1807

\bibitem[\protect\citeauthoryear{van der Marel et al.}{2001}]{vandermarel02}    
van der Marel R. P., Alves D. R., Hardy E., Suntzeff N. B., 2002, AJ, 124, 2639

\bibitem[\protect\citeauthoryear{Walker}{2012}]{w12}  
Walker A., 2012, Ap\&SS.tmp, 746 in press, arXiv:1112.3171


\end{thebibliography}
\end{document}